\documentclass{jfm}
\usepackage{graphicx}
\usepackage{epstopdf, epsfig}
\usepackage{amsmath} 

\shorttitle{Detonations with flow divergence}
\shortauthor{M. I. Radulescu and B. Borzou}

\title{Dynamics of detonations with a constant mean flow divergence}

\author{Matei I. Radulescu
      \and Bijan Borzou
      \corresp{\email{matei@uottawa.ca}}}

\affiliation{Department of Mechanical Engineering, University of Ottawa, ON K1N 6N5, Canada}

\begin{document}

\maketitle

\begin{abstract}
An exponential horn geometry is introduced in order to establish cellular detonations with a constant mean lateral mass divergence, propagating at quasi-steady speeds below the Chapman-Jouguet value. The experiments were conducted in 2C$_2$H$_2$+5O$_2$+21Ar and C$_3$H$_8$+5O$_2$.  Numerical simulations were also performed for weakly unstable cellular detonations to test the validity of the exponential horn geometry.  The experiments and simulations demonstrated that such quasi-steady state detonations can be realized, hence permitting to obtain the relations between the detonation speed and mean lateral flow divergence for cellular detonations in an unambiguous manner.  The experimentally obtained speed ($D$) dependencies on divergence ($K$) were compared with the predictions for steady detonations with lateral flow divergence obtained with the real thermo-chemical data of the mixtures.  For the 2C$_2$H$_2$+5O$_2$+21Ar system, reasonable agreement was found between the experiments and steady wave prediction, particularly for the critical divergence leading to failure.  Observations of the reaction zone structure in these detonations indicated that all the gas reacted very close to the front, as the transverse waves were reactive.  The experiments obtained in the much more unstable detonations in C$_3$H$_8$+5O$_2$ showed significant differences between the experimentally derived $D(K)$ curve and the prediction of steady wave propagation. The latter was found to significantly under-predict the detonability of cellular detonations.  The transverse waves in this mixture were found to be non-reactive, hence permitting to shed off non-reacted pockets, which burn via turbulent flames on their surface.  It is believed that the large differences between experiment and the inviscid model in this class of cellular structures is due to the importance of diffusive processes in the burn-out of the non-reacted pockets. The empirical tuning of a global one step chemical model to describe the macro-scale kinetics in cellular detonations revealed that the effective activation energy was lower by 14\% in 2C$_2$H$_2$+5O$_2$+21Ar and 54\% in the more unstable C$_3$H$_8$+5O$_2$ system.  This confirms previous observations that diffusive processes in highly unstable detonations are responsible for reducing the thermal ignition character of the gases processed by the detonation front. 
\end{abstract}


\section{Introduction}
Models for one-dimensional detonation waves in the presence of non-ideal effects (mass divergence, unsteadiness, friction, heat losses, turbulent fluctuations) rely on extensions of the classical one-dimensional Zeldovich-Von-Neumann-Doering (ZND) \citep{Fickett&Davis1979, Bdzil&Stewart2007, Lee2008}. In gases, real detonations are multi-dimensional and their reaction zone is often turbulent; deviations from the ZND model are substantial \citep{Lee2008,Shepherd2009} and cannot be treated as perturbations. Analysis of unstable detonations revealed that although the structure is highly transient, it recovers qualitatively a quasi-one-dimensional average structure analogous to the ZND structure, with an average sonic surface \citep{Gamezo1999, Radulescuetal2007, reynaud2017computational, maxwell2017influence}.  It thus appears that, with the appropriate kinetic description for the averaged fluid state and speed dictated by an effective rate of energy release, the ZND model with its extensions may be an appropriate framework to model real detonations at the macro-scale.  

The questions that arise are two-fold.  First, there is presently no direct way to extract the global kinetics of a real cellular detonation.  Secondly, it is of interest to establish whether the effective kinetics are faster or slower than those of the ZND model obtained from the underlying chemistry.  In other words, does the multi-dimensional cellular structure help or not the detonability of the mixture, as estimated from the ZND model and its extensions?  The present paper addresses these two questions.

The present study focuses on the dynamics of detonations with lateral mass divergence at the macro-scale.  We provide a novel geometry of an \textit{exponential horn}, which imposes a constant global lateral mass divergence to cellular detonations.  This permits to establish detonations in quasi-steady state and hence allows to make meaningful comparison with ZND models with lateral mass divergence.  It also allows for unambiguous determination of the global kinetic laws for cellular detonations. 

The theoretical problem of steady detonations with a constant lateral mass divergence is now well established \citep{He1994, Yao1995, Kleinetal1994}. In the presence of a flow divergence in the steady reaction zone, the divergence has the opposite effect to the thermicity, which accelerates the flow to the sonic condition.  For a given lateral flow divergence, or equivalently a front curvature ($\kappa$), there will be a unique detonation speed that can permit the post shock flow to accelerate to the sonic condition, where the thermicity must balance the lateral divergence. Given a $D(\kappa)$ relation for a given detonable mixture, the dynamics of detonations in complex geometries can be predicted by lower order models, for example in the framework of Detonation Shock Dynamics (DSD)  \citep{Bdzil&Stewart2007}. 

While the $D(\kappa)$ relation can be obtained from the real chemistry under the assumption of a steady detonation model \citep{Kleinetal1994}, it can also be extracted from experiments.  In condensed phase detonations, where the underlying chemical decomposition rates are generally not known with sufficient detail from first principles, this empirical approach is central to deduce the macro-scale kinetics of detonations \citep{Bdzil&Stewart2007}.  In gases, however, the kinetics are relatively well known.  The question that hence arises is whether $D(\kappa)$ curves obtained from the underlying chemistry from first principles agree with those characterizing the dynamics of cellular detonations with a global divergence.  

This question has been attempted in the past in a series of experimental and numerical studies.  Global mass divergence has been used to model detonations in narrow tubes, where the viscous boundary layers developing on the tube walls act as a mass sink to the core flow. The divergence of the streamlines in the reaction zone results in a globally curved detonation front experiencing a velocity deficit \citep{Dupreetal1991,Chinnayyaetal.2013, Moenetal1981}.  \citet{Chaoetal2009}, \citet{Camargoetal2010} and \citet{Gaoetal2016} attempted to compare the velocity deficits and limits observed in the experiments with predictions based on the ZND model with lateral mass divergence.  The amount of mass divergence was modelled globally using boundary layer theory using Fay's model \citep{Fay1959}.  The authors have found relatively good agreement for weakly unstable detonations, and poorer agreement for more unstable detonations. Nevertheless, there were a number of simplifying assumptions and matching constants, whose impact on the predictions have not been evaluated.  Firstly, Fay's model requires as empirical input a particular length scale for evaluating the mass divergence, typically taken as the induction length or the cell length of the non-attenuated detonation.  The model also assumes uniform curvature along the front, implying uniform area divergence for each stream tube in the reaction zone.  A curved detonation due to wall boundary layers or permeable walls is not expected to have a unique curvature, as demonstrated by \citet{Chinnayyaetal.2013} numerically.  Further away from the axis of the channel or tube, the flow divergence departs from that along the streamline along the axis.

\citet{Radulescu2003} and \citet{Radulescu&Lee2002} applied the concept of lateral mass divergence in the reaction zone of detonations to model detonation waves propagating in tubes with leaky, or porous walls. They conducted an extensive experimental series in several reactive mixtures and compared their results with the predictions of the ZND model with mass divergence.  In the models, the amount of mass divergence was assumed constant for all streamtubes, and evaluated from the porosity of the walls assuming choked flow.  Nevertheless, the assumption of constant divergence is also not realistic, as demonstrated numerically by \cite{Mazaheri2015} and argued above for detonations in narrow tubes and channels.  These limitations diminish the value of the comparison between experiment and ZND model predictions.  Nevertheless, interesting trends were observed.  For weakly unstable detonations, the model predictions using the real kinetics, particularly for the critical divergence rate, were found in good agreement with experiment.  For more unstable detonations, the agreement was much poorer.  The experimental detonations were found to propagate when the ZND models predicted extinction.  The authors interpreted this finding as the enhancement of the burning rate in more unstable detonations by multi-dimensional effects.  

Paradoxially and controversially, the role of cellular structures in numerically calculated inviscid detonations with mass divergence (solved with the Euler equations) has been shown to make the detonations more difficult to ignite than predicted from models derived from a laminar structure \citep{RadulescuetalICDERS2007, han2017role, han2017effect}.  Inviscid cellular detonations were found to be more prone to failure in weakly confined charges \citep{reynaud2017computational} and in the analogous problem of porous walled tubes \citep{Mazaheri2015}.  The authors found pockets of non-reacted gas that delay the global exothermicity.  These trends differed fundamentally from experimental observations, where the opposite trend was found, mainly that cellular detonations have an enhanced detonability than anticipated from ZND-type models neglecting the cellular structures, and that more unstable detonations were more detonable than stable ones.  It was recently suggested that turbulent mixing in the reaction zone of unstable detonations, absent in the inviscid simulations of the above studies, may account for these discrepancies \citep{maxwell2017influence}.  Indeed, diffusive effects were found to have profound differences even in one-dimensional detonation dynamics \citep{romick2012effect}.    

In the present study, we thus wish to clear this controversy and compare the $D(\kappa)$ curves for cellular detonations with those predicted from the underlying chemistry in a well posed experiment, which permits to establish whether the cellular structure of detonations increases or decreases the detonability of a given mixture.  

The current paper is organized as follows. Section 2 offers a brief justification of the usefulness of the \textit{exponential horn} geometry to study quasi-steady detonations with a constant global rate of mass divergence.  It also provides the mathematical problem solved to obtain the steady detonation solution with lateral flow divergence with real thermo-chemical data. Section 3 provides numerical simulations of weakly unstable detonations in an exponential horn.  These serve to evaluate the ability of an exponential horn to maintain multi-dimensional detonations in quasi-steady state. Section 4 provides the details of the experimental procedure and the experimental results obtained.  Section 5 provides a general discussion, which is followed by the paper's conclusion.   
 
\section{Quasi-1D flow with lateral mass-divergence}
\subsection{Reactive Euler equations for quasi 1D flow}
The exact equations of gasdynamics in multiple dimensions for a $N$-component system, which neglect transport terms, are the so-called reactive Euler equations.  The usual continuity, momentum conservation and energy conservation following a fluid particle are:
\begin{equation}
  -\frac{1}{\rho}\frac{D \rho}{D t} =\bnabla\bcdot\boldsymbol{u}
   \label{eq:REEQMASS}
\end{equation}
\begin{equation}
  \rho \frac{D \boldsymbol{u}}{D t}=-\bnabla p
   \label{eq:REEQMOM}
\end{equation}
\begin{equation}
  \frac{D e}{D t}=-\frac{p}{\rho^2}\frac{D \rho}{D t}
   \label{eq:REEQEN}
\end{equation}
where $D/Dt=\partial/\partial t+\boldsymbol{u}\bcdot\bnabla$ is the material derivative and the symbols have their usual meaning \citep{Williams1985}.  These need to be complemented by an appropriate equation of state for the internal energy $e$ of the form:
\begin{equation}
  e=e\left(p, \rho, Y_1, Y_2, ...,Y_i,..., Y_N \right) 
   \label{eq:REEeos}
\end{equation}
and the appropriate kinetics for the evolution of the mass fractions $Y_i$'s of each of the system's $N$ components: 
\begin{equation}
  \frac{D Y_i}{D t}= \frac{\omega_i}{\rho}
  \label{eq:REErates}
\end{equation}
When a quasi-1D formulation is sought in a duct (or streamtube) of cross-section $A(x)$ , the variables take the meaning of transverse averages (or streamtube variables) and the equations remain unchanged, with $D/Dt=\partial/\partial t+u\partial/\partial x$.  The divergence of the velocity field in the continuity equation (\ref{eq:REEQMASS}), which describes the rate of volume change of a fluid element per unit volume, can be expressed as two contributions.  
\begin{equation}
  \bnabla\bcdot\boldsymbol{u}= \frac{\partial u}{\partial x} + \dot{\sigma}_A
  \label{eq:div}
\end{equation}
The first is the usual rate of strain of a fluid element in the flow direction $x$, and the second, $\dot{\sigma}_A$, is the rate of strain in the transverse direction \citep{Whitham1974, Williams1985}: 
\begin{equation}
  \dot{\sigma}_A\equiv u \frac{d (lnA)}{dx} 
  \label{eq:thermicityA}
\end{equation}
This term plays a fundamental role in both inert \citep{Whitham1974} and reactive gasdynamics\citep{Fickett&Davis1979}, as it provides the rate of expansion or compression of a fluid element due to geometrical effects.  For such quasi-one-dimensional flows, the conservation laws are best cast in characteristic form, which, after some manipulation and basic thermodynamic identities \citep{Williams1985, Kao2008, Wood&Kirkwood1957}, yield:
\begin{equation}
  \frac{1}{\rho c^2} \frac{D_\pm p}{D t} \pm \frac{1}{c} \frac{D_\pm u}{D t}=\dot{\sigma} - \dot{\sigma}_A 
  \label{eq:characteristicform}
\end{equation}
where the derivatives 
\begin{equation}
\frac{D_\pm }{D t}= \frac{\partial}{\partial t}+(u \pm c)\frac{\partial}{\partial x}
  \label{eq:characteristics}
\end{equation}
are taken along the path of forward facing $C_+$ and rear facing $C_-$ characteristics, given respectively by $dx/dt=u \pm c$.  The thermicity $\dot{\sigma}$ denotes the gasdynamic effect of chemical reactions or other relaxation phenomena on the rate of pressure and speed changes along the family of characteristics.  We immediately remark that the effect of transverse flow divergence $\dot{\sigma}_A$ has precisely the opposite effect to the thermicity $\dot{\sigma}$ on modulating the strength of acoustic waves along the characteristic paths.   

The thermicity in its most general form is given by:
\begin{equation}
\dot{\sigma}= - \frac{\rho}{c_p} \left(\frac{\partial v}{\partial T}\right)_{p, Y_i} \sum_{i=1}^N \left( \frac{\partial h}{\partial Y_i}\right)_{p, \rho, Y_{j,j \neq i}}\frac{D Y_i}{D t}
  \label{eq:thermicity}
\end{equation}
\citep{Fickett&Davis1979, Williams1985, Kao2008}.  For a mixture of ideal gases, the thermicity reduces to
\begin{equation}
\dot{\sigma}= \sum\limits_{i=1}^{N}\left(\frac{\bar{W}}{W_i}-\frac{h_i}{c_pT}\right) \frac{D Y_i}{D t}
  \label{eq:thermicityidealmix}
\end{equation}
\noindent where $W_i$ is the molecular weight of the i$^\text{th}$ component, $\bar{W}$ is the mean molecular weight of the mixture, $h_i$ is the specific enthalpy of the i$^\text{th}$ specie and $c_p$ is the mixture frozen specific heat.  This expression takes a simple and more familiar form for a reacting perfect gas with an equation of state of the form 
\begin{equation}
p=\rho R T
\end{equation}
\begin{equation}
   e(p, \rho, Y_R, Y_P)=\frac{1}{\gamma -1}\frac{p}{\rho} + \left(1-Y_P\right)Q
  \label{eq:eosperfect}
\end{equation}
where $Y_R$ and $Y_P$ are the mass fractions of the reactants and products, respectively, $\gamma$ is the isentropic exponent and $Q$ is the chemical energy released during the reaction \citep{Fickett&Davis1979}:  
\begin{equation}
   \dot{\sigma}=  \frac{(\gamma-1) \rho Q}{\gamma p} \frac{D Y_P}{D t}
  \label{eq:thermicityperfect}
\end{equation}

In the present paper, we will deal with flows for which the source term appearing in the quasi-one-dimensional formulation for a streamtube or in a tube of cross-sectional area $A(x)$, namely:
\begin{equation}
  K\equiv \frac{d (lnA)}{dx}
 \label{eq:kappadef}
\end{equation}
is a \textit{constant}.  The source term being a constant permits us to establish a steady (on average) detonation structure, as can be anticipated since the source is neither a function of space nor time.  In the current work, we study the dynamics of detonations in a channel with a constant logarithmic derivative $K$, i.e., a channel with a cross-section varying exponentially with distance. This is known as an \textit{exponential horn} in acoustics, since, in the absence of chemical reactions, acoustic waves will have constant amplification factors by virtue of (\ref{eq:characteristicform}). 

\subsection{Steady wave solution}
The steady traveling wave solution in the presence of lateral mass divergence for quasi 1D flow is obtained in the usual way \citep{Wood&Kirkwood1954, Kleinetal1994, Yao1995, Bdzil&Stewart2007, He1994, Kasimov&Stewart2005, Short&Bdzil2003}.  Typically, the flow divergence in the detonation frame of reference is kept constant.  In our experiments, however, the logarithmic derivative of the stream tube $A$ in the laboratory frame of reference, given by (\ref{eq:kappadef}) is kept constant, to reflect the exponential horn geometry.  

Note that the logarithmic derivative of the streamtube cross-sectional area is also the curvature of the detonation front $\kappa$ \citep{Fickett&Davis1979}:
\begin{equation}
  \kappa = K = \frac{d (lnA)}{dx}
 \label{eq:curvaturedef}
\end{equation}
\noindent such that we seek a travelling wave solution that maintains a constant front curvature.  

A physically relevant derivation of the steady wave solution starts with the governing equations for quasi-1D gasdynamics written in characteristic form (\ref{eq:characteristicform}).  Seeking a traveling wave solution with speed $D$, we change to a wave-fixed frame with the change of variables $x'=D t -x$, $t'=t$ and $v=D-u$ and requiring that the unsteady terms vanish (for the steady traveling wave), the differential operators transform as
\begin{eqnarray}
   \frac{D_\pm}{Dt}=\left(v\mp c\right) \frac{d}{dx'}\\
   \frac{D}{Dt}=v\frac{d}{dx'}
\end{eqnarray}
With this transformation, the characteristic equation (\ref{eq:characteristics}) becomes:
\begin{equation}
  \frac{1}{\rho c^2} \frac{dp}{dx'} \mp \frac{1}{c} \frac{dv}{dx'}
  =\frac{\dot{\sigma} - \dot{\sigma}_A}{v \mp c}
  \label{eq:characteristicformsteady}
\end{equation}
with the upper and lower signs for the $C_+$ and $C_-$ families of characteristics, respectively. Solving for the derivatives, we obtain the usual relations for the steady ZND structure of detonations:
\begin{equation}
  \frac{dv}{dx'}=-\frac{1}{\rho v}\frac{dp}{dx'}=\frac{\dot{\sigma} -\dot{\sigma}_A}{1 - M^2} 
  \label{eq:ZNDrelationspandu}
\end{equation}
with a similar expression for the density derivative:
\begin{equation}
  \frac{d \rho}{dx'}=-\frac{\rho}{v} \left(\frac{dv}{dx'} +\dot{\sigma}_A \right)
  \label{eq:ZNDrelationsden}
\end{equation}
\noindent and the evolution of the chemical composition of the mixture:
\begin{equation}
  \frac{d Y_i}{dx'}=-\frac{\omega_i}{\rho v} 
  \label{eq:ZNDrelationsspecies}
\end{equation}
With the appropriate kinetics to close the thermicity via (\ref{eq:thermicity}), (\ref{eq:thermicityidealmix}) or (\ref{eq:thermicityperfect}), the structure of the travelling wave solution is readily obtained by integrating (\ref{eq:ZNDrelationspandu}-\ref{eq:ZNDrelationsspecies}) from the shock at $x'=0$.  For a given shock speed $D$, the post shock state where the integration starts is given by the shock jump relations.  The flow begins subsonic, but tends to increase its speed $v$ due to energy release (positive thermicity).  The lateral flow divergence has the opposite effect.  As the flow may eventually become sonic, hence gasdynamically isolating the reaction zone from the rear flow, singular behaviour (marking non-steady dynamics) in (\ref{eq:characteristicformsteady}), (\ref{eq:ZNDrelationsden}) and (\ref{eq:ZNDrelationspandu}) is only prevented by the local balance between the thermicity $\dot{\sigma}$ and the rate of lateral strain $\dot{\sigma}_A$.  This is the so-called generalized Chapman-Jouguet criterion.  This condition provides the rear boundary condition.  The particular values of detonation speed that yield post-shock flows which will satisfy the generalized CJ condition somewhere in the reaction zone behind the lead shock will mark the possible steady state travelling wave solutions. Note that an interesting interpretation of this criterion can be physically interpreted by invoking the characteristic structure given by (\ref{eq:characteristicformsteady}): the reaction zone structure can be seen to consist of a region where forward facing characteristics communicate pressure wave amplification forward to sustain the lead shock, while the limiting characteristic travels at precisely the lead shock speed, along which the vanishing of the overall thermicity ensures that the waves travelling forward are not amplified. 

The structure of the traveling wave solution and the dependence between lateral flow divergence rate $K$ and detonation speed has been obtained in the present study for a one step perfect reacting gas, and also for a mixture of ideal gases with realistic thermal and chemical kinetic properties.  For the simple one step model, the ordinary differential equations were integrated using the numerical ODE solvers provided by Mathematica.  For the real gas calculations, a custom Python code was developed.  It uses Cantera's framework to evaluate the relevant thermo-chemical data \citep{Cantera} and the 
Shock and Detonation Toolbox for evaluating the shock jump conditions \citep{KaoShepherd2008}.  The resulting system of ODE's were integrated using the VODE stiff solver provided by SciPy. The shooting procedure to obtain the steady detonation structure with lateral mass divergence follows the method outlined by \citep{Kleinetal1994}.  Our code was verified by comparing the results with those provided by \citep{Kleinetal1994} using their model for the stream tube area variation.  

\begin{figure}
  \centerline{\includegraphics[width=0.7\textwidth]{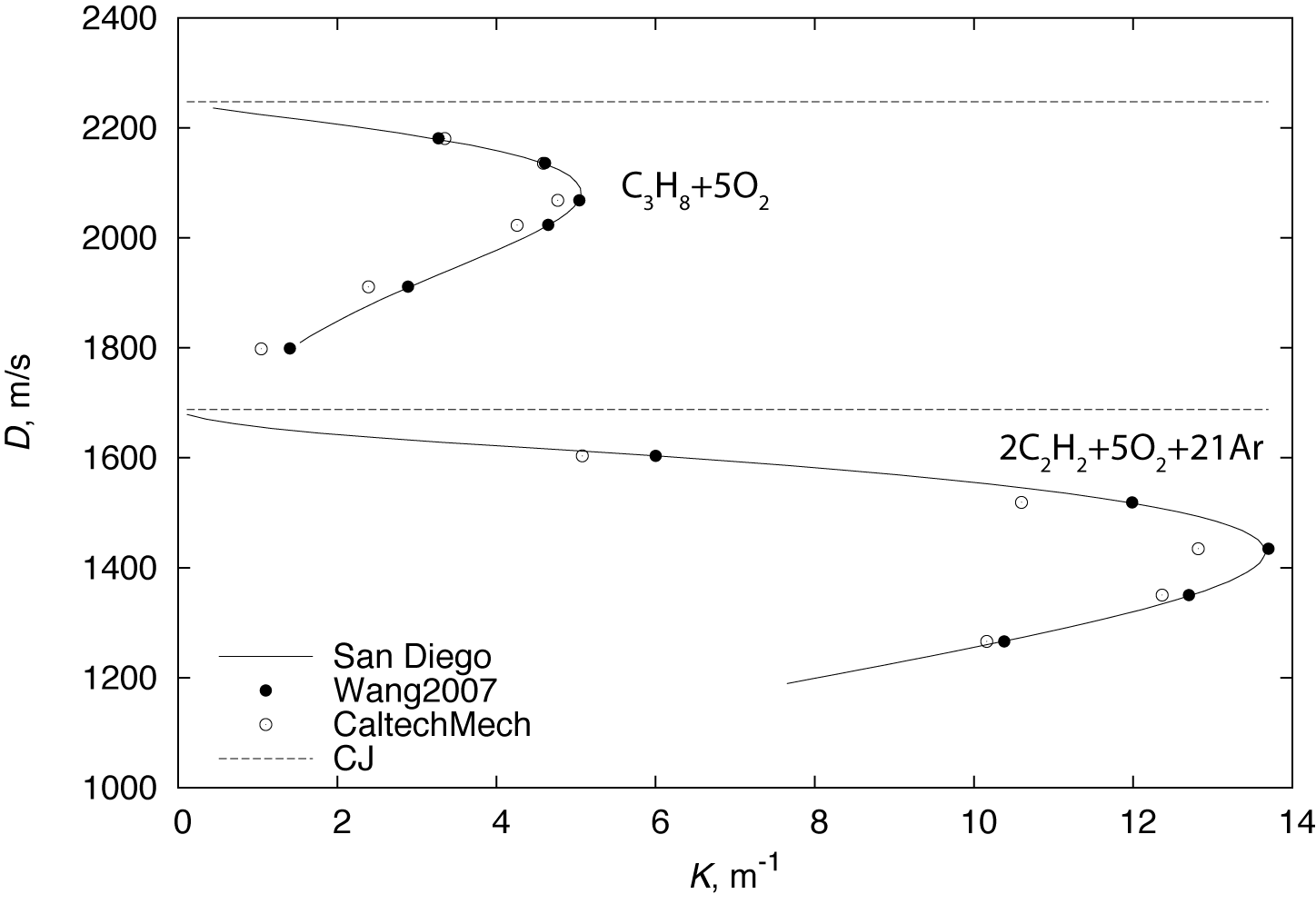}}
  \caption{Relation between the detonation speed and lateral flow divergence rate $K$ for 2C$_2$H$_2$+5O$_2$+21Ar and C$_3$H$_8$+5O$_2$ at 8.6 kPa initial pressure obtained with different chemical kinetic mechanisms; dotted lines represent the ideal Chapman-Jouguet (CJ) speed.}
\label{fig:dkexample}
\end{figure}

Figure \ref{fig:dkexample} shows, for example, the numerically obtained $D(K)$ curves for the two mixtures investigated in the present study, namely 2C$_2$H$_2$+5O$_2$+21Ar and C$_3$H$_8$+5O$_2$ at 8.6 kPa initial pressure using the San Diego chemical mechanism \citep{SanDiego}. Results are also provided using two other chemical kinetic mechanisms widely used in the literature and also validated against published ignition delay data for these mixtures, the Wang2007 mechanism \citep{wang2007} and the CaltechMech mechanism \citep{CaltechMech}. The $D(K)$ curves show minimal influence on which kinetic mechanism was used and display the characteristic decay of detonation speed with increasing lateral flow divergence, as well as the turning point denoting the maximum permissible lateral flow divergence. In this paper, further calculations will use the San Diego chemical mechanism.  


\section{Validation of quasi 1D assumption for the exponential horn geometry}\label{sec:exponential horn}
In the present study, we propose using an exponential horn, such that the cross-section area increases at a rate to maintain its logarithmic derivative $K$ a constant.  Since the quasi-1D approximation is only expected to hold in the limit of small divergence, it is worthwhile to study the departures from the quasi-1D approximation for two-dimensional detonations propagating in an exponential horn.  We treat this problem numerically for gas parameters yielding a very weak cellular structure, such that the cellular structure does not strongly influence the dynamics. We document the magnitude of higher order effects in the dynamics that do not permit 2D detonations to be approximated as quasi-1D. 

In this section, we report the results of two-dimensional unsteady calculations of detonations propagating in a diverging channel. For simplicity, we consider the standard single step Arrhenius reaction in a perfect gas, for which the rate of depletion of reactant is related to the rate of production of product by 
\begin{equation}
  \omega_P=-\omega_R=k \rho \left(1-Y_P \right) e^{-\frac{E_a}{RT}}
  \label{eq:onestep}
\end{equation}
\noindent We use the initial pressure $p_0$, initial density $\rho_0$ and the detonation half reaction length $\Delta_{1/2}$ as dimensional quantities to non-dimensionalize the problem; the resulting energy scale is thus $p_0/\rho_0$, velocity scale is $\left(p_0/\rho_0\right)^{1/2}$ and time scale is $\Delta_{1/2}\left(p_0/\rho_0\right)^{-1/2}$.  Below, symbols with an over-bar refer to dimensionless variables normalized in this manner.  The heat release $\bar{Q} = 11.5$ and specific heat ratio $\gamma= 1.54$ correspond approximately to a mixture of 2H$_2$+O$_2$+7Ar, while the effective activation energy $\bar{E_a} = 15$ is taken significantly lower than the real one in order to obtain a very weak cellular structure \citep{sharpe2008nonlinear}.   

\begin{figure}
  \centerline{\includegraphics[width=.5\textwidth]{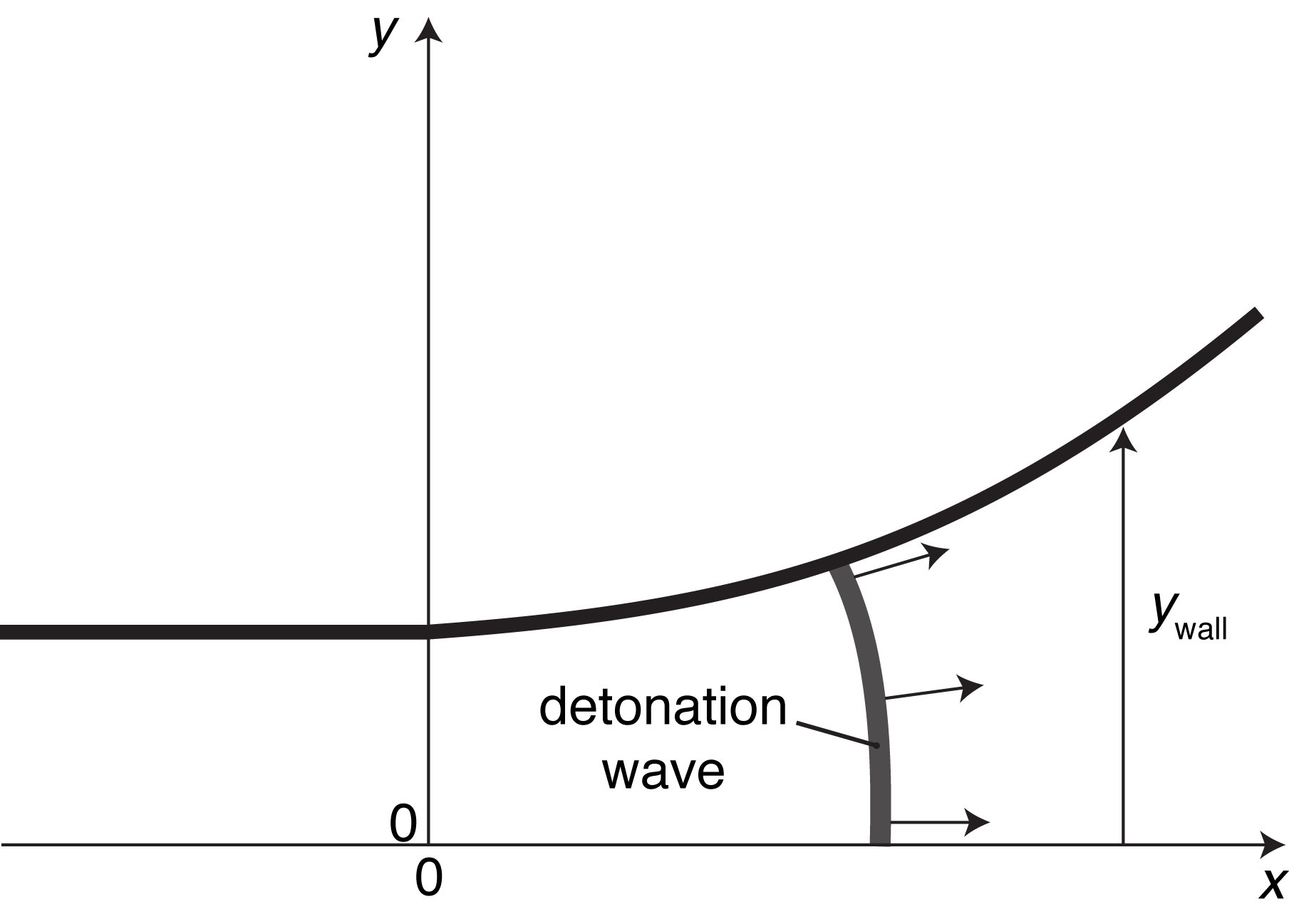}}
  \caption{Definition of the problem solved numerically.}
\label{fig:numericalproblem}
\end{figure}

The initial boundary value problem solved is the evolution of an inviscid detonation wave, obeying (\ref{eq:REEQMASS}-\ref{eq:REEQEN}, \ref{eq:eosperfect}), which consumes a perfect gas whose depletion is given by (\ref{eq:onestep}), activated by the arrival of a lead shock.  The initial conditions correspond to a ZND detonation, whose lead shock is initially at $x=0$ and whose reaction zone extends to $x \rightarrow -\infty$.  Ahead of the detonation wave, the quiescent state is $p=\rho=1$ and $u=v=0$.  The channel's wall is given by $y_{wall}=y_0e^{Kx}$ for $x>0$ and by $y_{wall}=y_0$ otherwise.  Along the wall, a no penetration boundary condition applies while the axis $x=0$ has a symmetry boundary condition.   Figure \ref{fig:numericalproblem} illustrates the problem.

The dynamics of the detonation wave were investigated in the diverging sections for divergence rates $\bar{K}$ of 0.004 and 0.008.  The former represents the largest typical value in the experiments, while the latter is used for studying the ensuing errors at large divergences for illustrative purposes.

The numerical solution to the initial boundary value problem was obtained using the MG code developed by S. Falle of the University of Leeds. The convective terms in the governing equations were evaluated using an exact Riemann solver \citep{richtmyer1967difference, Falle1991}, which features a symmetric monotonized central flux limiter \citep{vanleer1977towards}. The time evolution of the chemical source terms was performed explicitly and coupled to the hydrodynamics by the method of fractional time steps. Structured Cartesian grids are used to take advantage of adaptive mesh refinement (AMR) \citep{falle1993body}.  AMR is controlled by user defined differences between the solutions computed at different grid levels. In this regard, a hierarchical series of grids $G^0 ...G^n$ are used, so that grid $G^n$ has the mesh spacing $h/2^n$ where $h$ is the coarse grid size. The advanced solution on the grids $G^{n-1}$ and $G^n$ are compared on a cell by cell basis, to decide whether cells on the latter need to be refined.  

In the simulations, the base grid spacing $h$ was set to 2 $\Delta_{1/2}$.  7 grid levels ($n$ = 6) were used in order to attain a minimum grid spacing of $\delta$=1/32 $\Delta_{1/2}$, i.e. a resolution of 32 grid points per half reaction zone length of the steady CJ wave.  The finest grids always covered the reaction zone and the curved wall.  The resolution study presented in the appendix showed that neither the cellular structure nor the mean detonation speed were affected for these very weakly unstable detonations, yielding an error on the detonation speeds reported less than 0.002. 

The numerical calculations were evolved on the grid fixed on the "laboratory" frame of reference.  In this frame of reference, the gas ahead of the detonation wave is at the initial state and its velocity is zero for the entire calculation, which ends when the detonation wave front reaches the right boundary.  For this reason, the right boundary condition is not relevant.   The height of the numerical domain was always 100$\Delta_{1/2}$. The length of the domain was such as to accomodate the desired exponential horn geometry.  The rear of the domain extended to -100.  The curved wall was implemented directly on the Cartesian grid in a staircase fashion, using a symmetry boundary condition on each cell surface representing a gas-solid interface. This appeared as adequate when a sufficiently high resolution was used, as evidenced by the convergence study of the appendix.  The axis $y=0$ used a symmetry boundary condition.  The rear boundary condition used the fixed CJ post-detonation state. It was verified that the numerical domain extended sufficiently in the $-x$ direction such that right running waves from rear left boundary did not reach the detonation reaction front during the calculation. 
 
\begin{figure}
  \centerline{\includegraphics[width=1\textwidth]{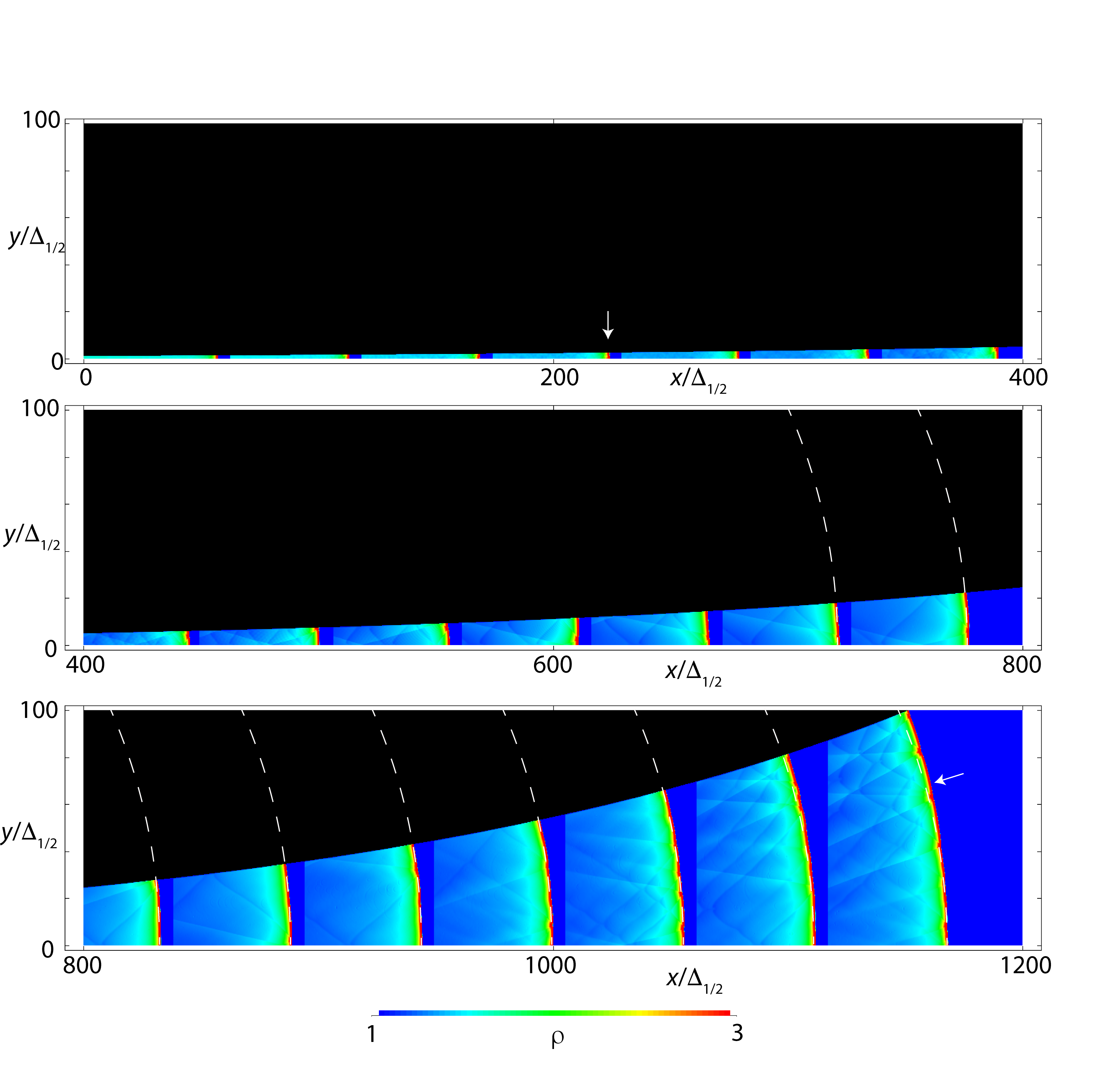}}
  \caption{Evolution of a curved detonation in an exponential horn with $K \Delta_{1/2}$=0.004; white lines denote arcs of circle with the expected curvature; arrows denote the two frames used for evaluating the mean speed of the wave along the bottom and top walls.}
\label{fig:K004_stable}
\end{figure}

Figure \ref{fig:K004_stable} shows the detonation evolution for $\bar{K} =$ 0.004.  The detonation was initiated in a channel of unity cross-section in this case.  A weak transverse wave structure develops on the detonation front, which also acquires a characteristic curvature.  The dashed white lines are arc of circles of the expected radius of curvature of the front of 1/0.004=250.  The front curvature is found in very good agreement with the curvature expected by the quasi-1D approximation used to design the exponential ramp.  Some minor deviations are observed in the last frame.

\begin{figure}
  \centerline{\includegraphics[width=0.7\textwidth]{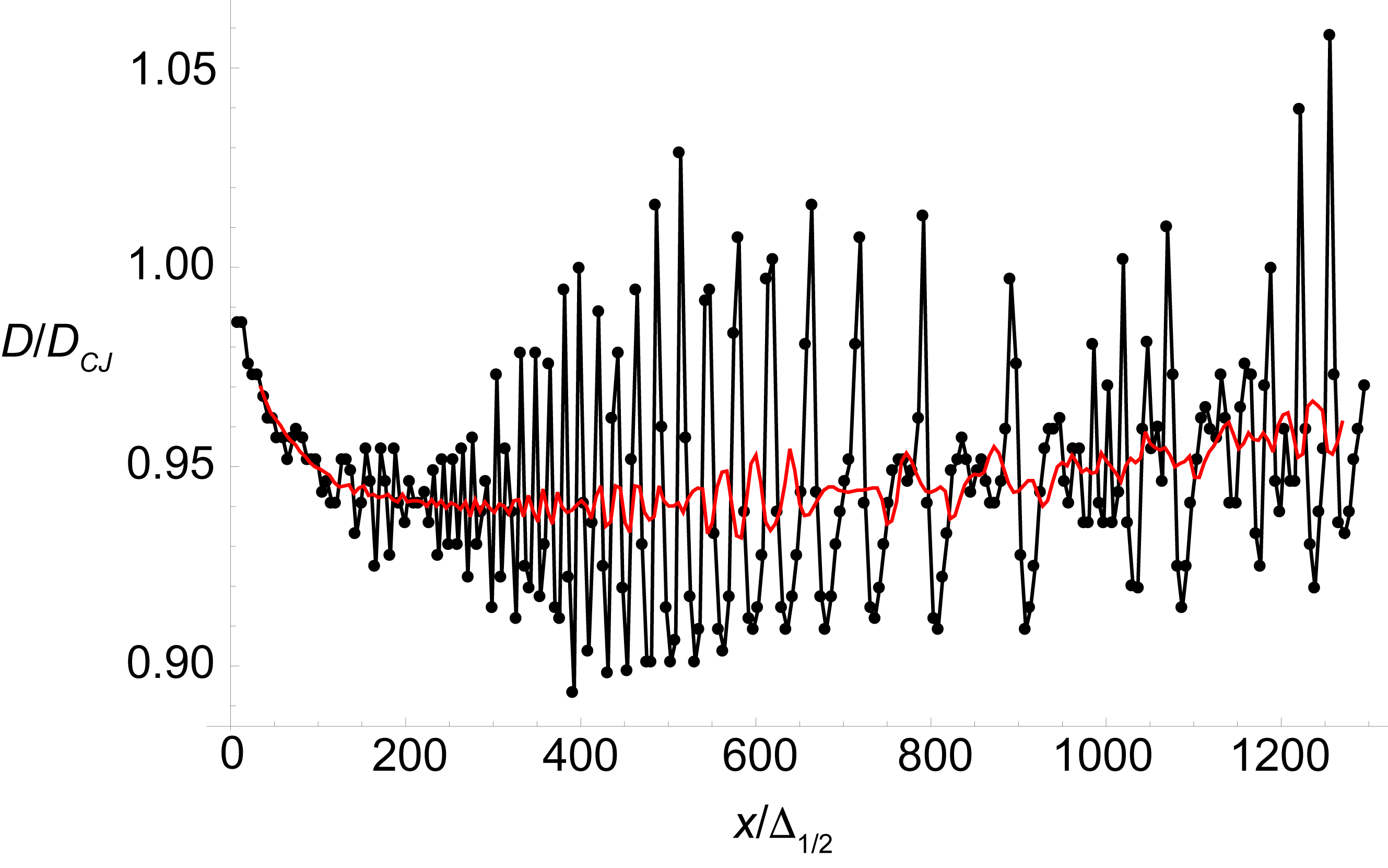}}
  \caption{Speed of the detonation illustrated in Fig. \ref{fig:K004_stable} measured along the bottom wall and its local time-average.}
\label{fig:K004_stable_speeds}
\end{figure} 

Figure \ref{fig:K004_stable_speeds} shows the evolution of the detonation front speed recorded along the bottom wall of the channel, as well as a running time average of period 10.  The cellular structure represents local deviations of speed of approximately 10\%.  The time average signal shows that the detonation mean speed is very nearly constant for the entire travel, although an apparent 1\% increase is also evident.  This deviation from the expected constant speed represents the limitation of the quasi-1D assumption in the design of the exponential horn.  

It is also instructive to compare the detonation speed along the straight and curved walls.  For an exponential wall given by $y_{wall}=y_0 e^{K x}$, the arc length $S$ of wall segment between two points of coordinates $(x_1, y_{wall}(x_1)$ and $(x_2, y_{wall}(x_2)$ is
\begin{equation}
S=\frac{1}{K}\left[\sqrt{1+(K y_{wall}(x))^2} - \tanh^{-1}\left(\sqrt{1+(K y_{wall}(x))^2}\right)\right]_1^2
\end{equation}
 Taking the interval between the two frames indicated by arrows in Fig. \ref{fig:K004_stable}, we obtain a mean speed $D/D_{CJ}$ along the bottom wall of 0.945 and along the curved wall of 0.939.  The slight difference is again attributable to the departures from the quasi-1D approximation.  In conclusion, the exponential horn performs as expected in maintaining a detonation with a constant mean lateral divergence rate and constant mean speed. The exponential horn also tends to provide an expansion that is sub-exponential, the correction originating from finite curvature effects of the multi-dimensional front.
       
\begin{figure}
  \centerline{\includegraphics[width=1\textwidth]{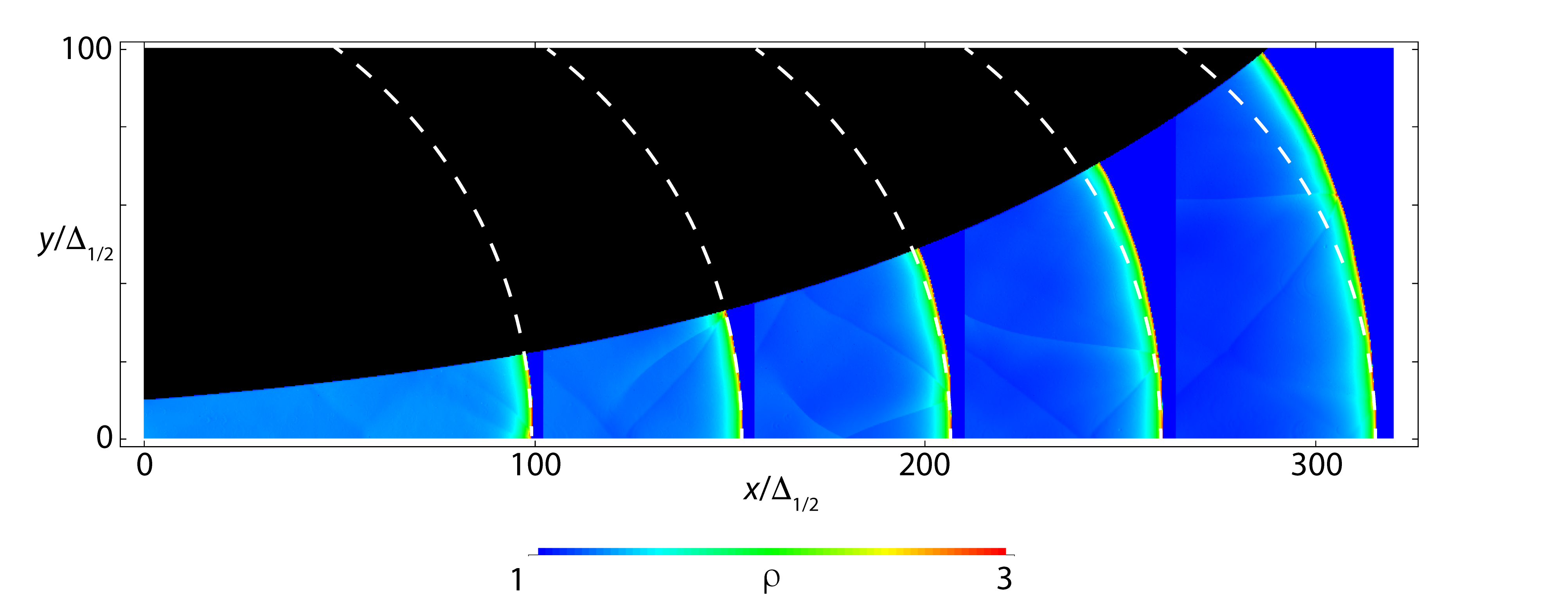}}
  \caption{Evolution of a curved detonation in an exponential horn with $K \Delta_i$=0.008; white lines denote arcs of circle with the expected curvature.}
\label{fig:K008_stable}
\end{figure}

Figure \ref{fig:K008_stable} shows the detonation evolution for $\bar{K}=$ 0.008.  The detonation was initiated in a channel of cross-section of 10 and allowed to travel until the cross-section increases to 100.  The larger initial cross-section was chosen to study the entrance transients.  For this larger divergence rate, which means that detonation run time was shorter, only a very weak transverse wave structure develops on the detonation front, which also acquires its characteristic curvature.  The dashed white lines denoting arc of circles of the expected radius of curvature of the front of 1/0.008=125 are found to approximate well the frontal curvature until the front reaches approximatelt $\bar{y} \approx 50$, with noticeable departures in wider channels. 

\begin{figure}
  \centerline{\includegraphics[width=0.7\textwidth]{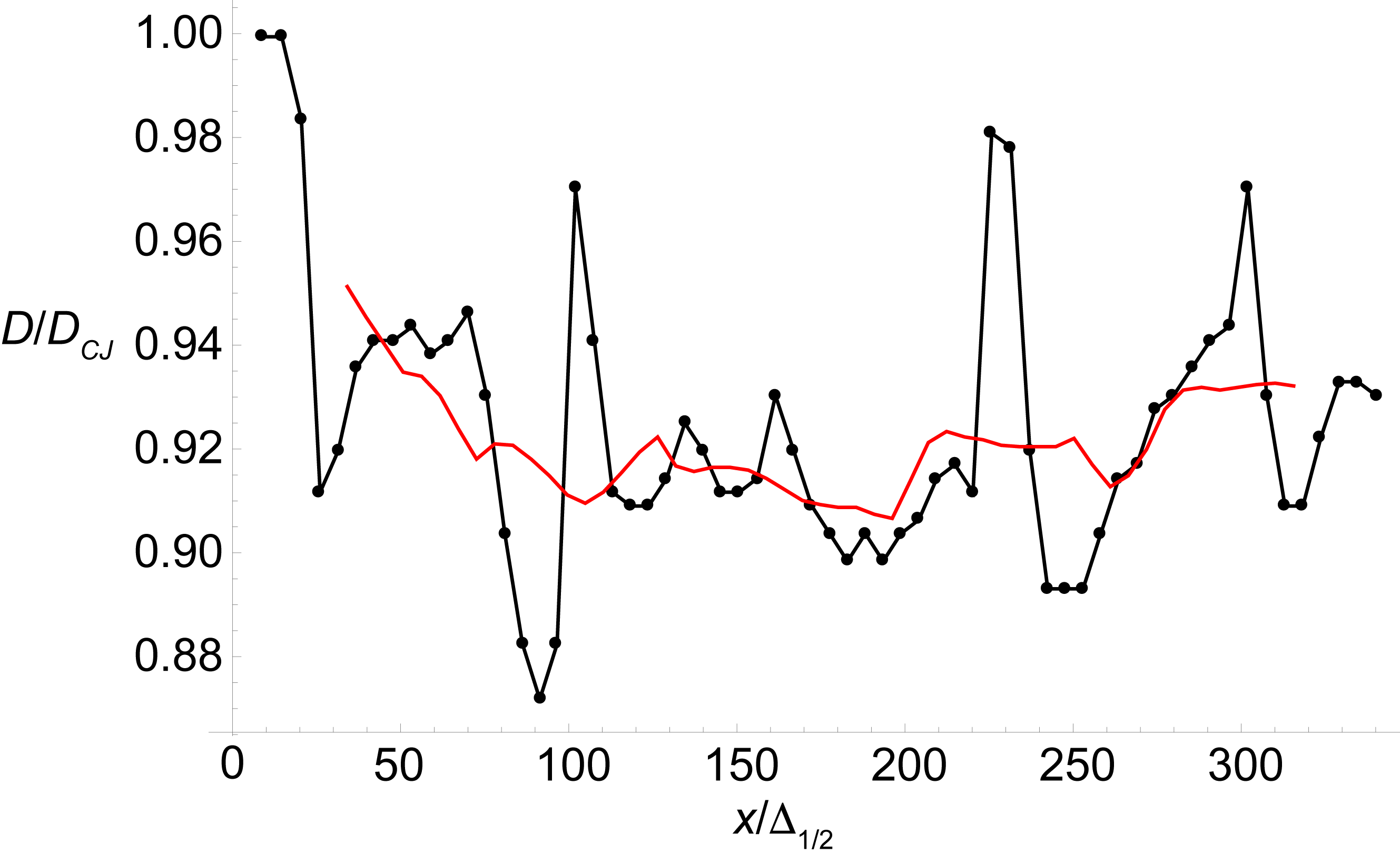}}
  \caption{Speed of the detonation illustrated in Fig. \ref{fig:K008_stable} measured along the bottom wall and its local time-average.}
\label{fig:K008_stable_speeds}
\end{figure}   

Figure \ref{fig:K008_stable_speeds} shows the evolution of the detonation front speed recorded along the bottom wall of the channel, as well as a running time average of period 10.  A quasi-steady evolution is evident, again with a slight increase of approximately 2\% over the length of travel.  Entrance transients appear to have decayed after approximately 10 entrance heights.  For this more curved detonation, the speed of the wave along the bottom wall, measured between the first and last frame showed in Fig. \ref{fig:K008_stable} was found to be approximetely 0.92, while the detonation speed along the curved wall was 0.88.  The larger difference highlights the limitations of the quasi-1D approximation at larger divergence rates in large channels.  

The error in the quasi-1D approximation can be estimated by comparing the effective curvature of the lead shock wave with the expected one from the quasi 1D approximation.  In an exponential horn characterized by its constant $K$, the arc of circle intersecting the wall at $(x, y_{wall}(x)$ perpendicularly has a curvature $K_{2D}$ given by
\begin{equation}
\frac{K_{2D}}{K} = \left( 1 + \left( K y_{wall} \right)^2 \right)^{-1/2}\approx 1 - \frac{1}{2} \left( K y_{wall} \right)^2
\end{equation}
The last term thus represents an estimate for the error in the quasi-1D approximation.  For example, for the $K \Delta_{1/2}$ of 0.008 and a mean square averaged channel dimension of 71, the estimated error in the effective lateral divergence rate is 16\%.  For the $K \Delta_{1/2}$ of 0.004, the error is 4\%.

\begin{figure}
  \centerline{\includegraphics[width=0.7\textwidth]{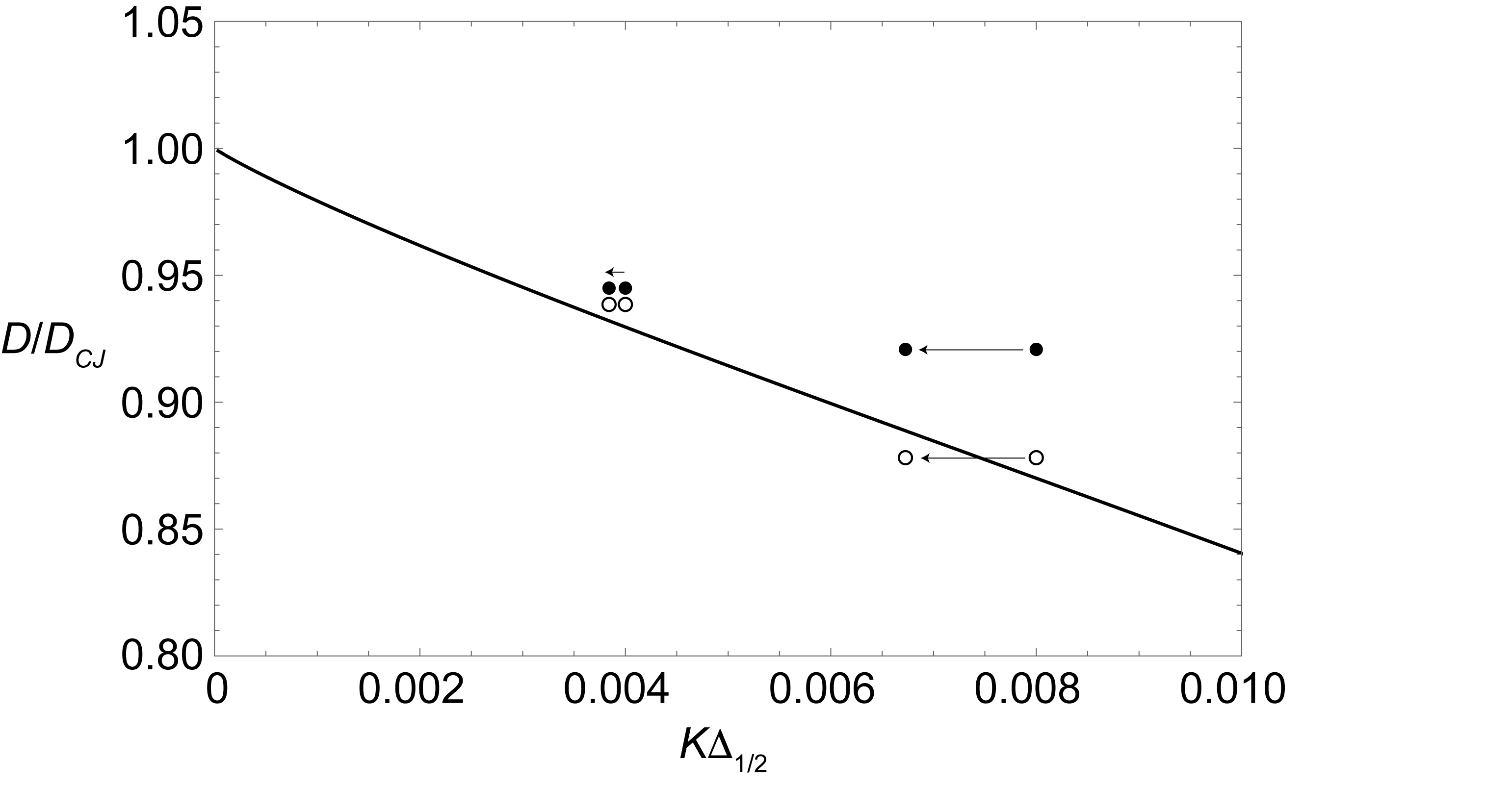}}
  \caption{Detonation front speed measured along the top wall (open symbols) and along the bottom wall (closed symbols); solid line is the quasi-1D ZND prediction; arrows indicate the corrections for the 1D approximation in an exponential horn.}
\label{fig:stable_speeds}
\end{figure}

It is also of interest to compare the wave speeds recorded in the simulations with those predicted from the quasi-1D ZND model with lateral mass divergence discussed in the previous section. Figure \ref{fig:stable_speeds} shows this comparison.  While the detonation speeds recorded from the simulations are found larger than predicted, the difference is reduced when correction for the effective divergence rates are effected, denoted by the arrows in Fig. \ref{fig:stable_speeds}.  Nevertheless, the detonations are found approximately 1-2 \% faster than predicted by the quasi-1D formalism.  It is unclear at present what causes this small departure and whether it is associated with the weak cellular structure or a longer entrance transient.  

The overall conclusion of the present section is that the exponential horn geometry is able to maintain a detonation in quasi-steady conditions with a mean lateral flow divergence that is kept constant.  Departures from the constant mean divergence and steady speeds are weak, with a tendency for the effective divergence rate to be under-estimated, as the front is mildly accelerating.  The technique thus appears adequate to determine the $D(K)$ curves experimentally for cellular detonations for moderate divergence rates.  This is addressed in the next section.  

\section{Experiments}\label{sec:Experiments}
\subsection{Experimental setup}

\begin{figure}
  \centerline{\includegraphics[width=1\textwidth]{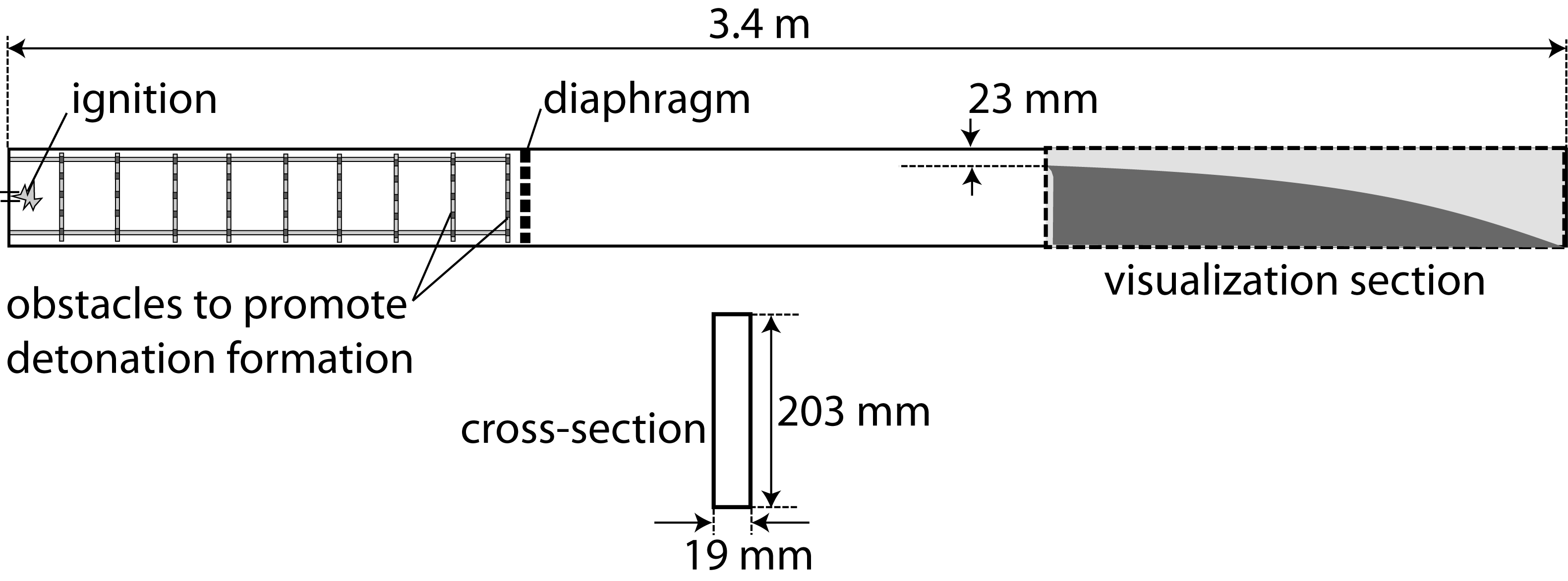}}
  \caption{A schematic of the experimental apparatus to study diverging detonations.}
\label{fig:Shktube}
\end{figure}

Figure \ref{fig:Shktube} shows the schematic of the 3.4-m-long rectangular channel used for the experiments in this paper. The channel is made of aluminium. The internal height and width of the channel are 203 mm and 19 mm respectively. This internal width was found to be the optimum value to suppress the transverse perturbations and thus approximate the detonation cellular dynamics as two-dimensional \citep{Bhattacharjee2013}. 

The channel had three identical sections (Fig. \ref{fig:Shktube}). The diverging ramp was placed in the third section, which was equipped with glass panels allowing to visualize the detonation front evolution.  Two exponentially diverging ramps were used for the tests, which permitted to correct for boundary layer losses in a thin channel geometry, as described later.  A long ramp (with 1-m-length and $K=\frac{1}{A}\frac{dA}{dx}$=2.302  m$^{-1}$) and a short ramp (with 0.5-m-length and $K=\frac{1}{A}\frac{dA}{dx}$= 4.605 m$^{-1}$) were used.  At the entrance of the diverging section, the initial opening was 23 mm in height. A "cookie-cutter" entrance was used in order to isolate the detonation front entering the diverging section. The diverging ramp was machined from a sheet of polyvinyl-chloride (PVC).  

The detonation was initiated by two different methods in the first section of the channel.  In the first method, a powerful spark from capacitor discharge was used.  Its discharge characteristics are described in detail by \citet{Bhattacharjee2013}. The capacitor bank nominally stored approximately 1000 J at a charging voltage of approximately $V_0$=32 kV.  However, a useful energy of only approximately 5-10 J was discharged in approximately 5 $\mu$s.  This generated a strong blast wave giving rise to a high-speed deflagration wave. A wire-mesh grid was installed in the first section of the tube in order to ensure the rapid transition of the deflagration wave to a detonation.  This method, however, was not successful in the experiments conducted with the acetylene mixture at initial pressures below 8 kPa.  For these conditions, a more sensitive reactive driver was used in the first section of the channel, which was separated from the test mixture by a thin Mylar diaphragm (see Fig. \ref{fig:Shktube}).  The driver gas used was C$_2$H$_4$+3O$_2$. The driver gas was ignited using the same capacitor discharge as described above, in which a detonation was established. This incident detonation wave was found to transmit successfully into the test mixture.   

The mixtures tested were 2C$_2$H$_2$+5O$_2$+21Ar and C$_3$H$_8$+5O$_2$.  The mixtures were prepared by the method of partial pressures in a separate mixing tank.  The mixtures were typically left to mix over a period of at least 24 hours prior to an experiment.  Before an experiment, the channel was first evacuated to a pressure below 100 Pa prior to the injection of the test mixture.  The experiments were performed at ambient room temperature of 20 $\pm$ 2$^{\circ}$C.  The sensitivity of the test mixtures was controlled by varying the initial pressure of the gases.  

The visualization of the detonation wave front evolution in the diverging channels was performed using a large-scale shadowgraph system described by \citet{Dennis2013}. A high speed Phantom V1210 camera was used for visualization.  Images were recorded at resolutions of 1152 by 256 and a framing rate of approximately 42000 frames per second, yielding a time resolution of approximately 24 $\mu$s.  An exposure of 1 $\mu$s was used. For reference, typically 20 consecutive frames were acquired to monitor the detonation passage through the test section.  These frames were then post-processed in order to evaluate the speed of the lead shock between successive frames.  These were recorded along the top wall only.  The typical errors in evaluating these speeds, derived from the image resolution and the ability to identify the lead shock position in the photographs was approximately $\pm 2\%$ \citep{Borzou2016}.   

\subsection{Results in 2C$_2$H$_2$+5O$_2$+21Ar}
\begin{figure}
  \centerline{\includegraphics[width=1\textwidth]{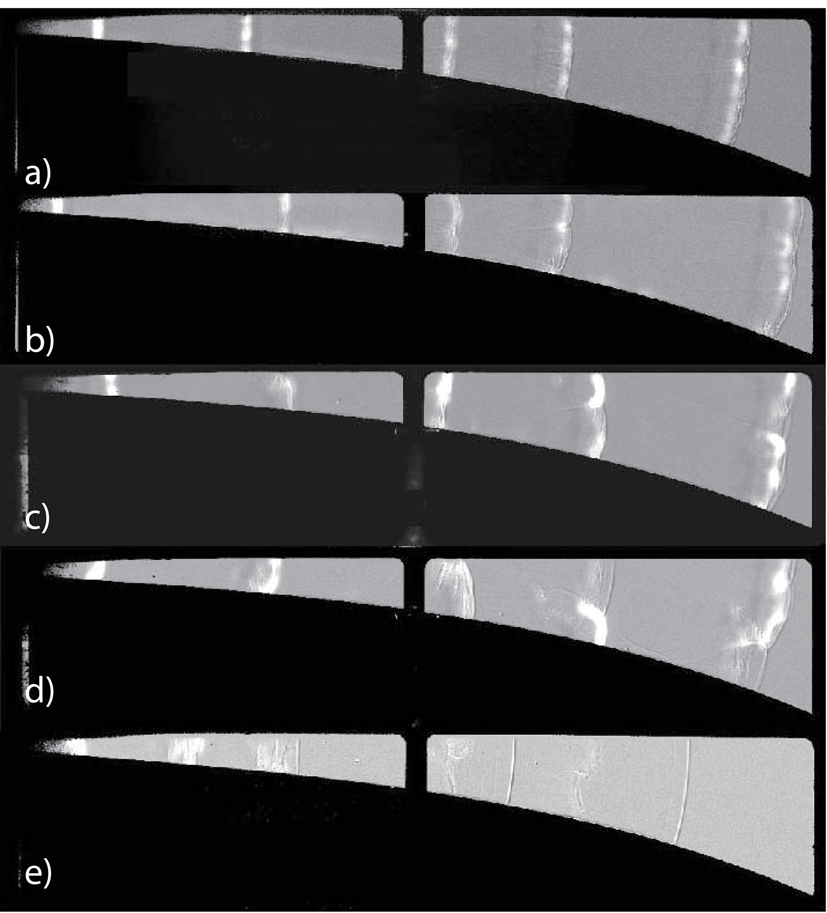}}
  \caption{Detonation front evolution in 2C$_2$H$_2$+5O$_2$+21Ar along the long ramp at initial pressures of (a) 12.1 kPa, (b) 9 kPa, (c) 6.2 kPa, (d) 4.9 kPa and (e) 4.2 kPa.}
\label{fig:AL1}
\end{figure}

Figure \ref{fig:AL1} shows the evolution of the detonation wave propagation for five different experiments in 2C$_2$H$_2$+5O$_2$+21Ar at different initial pressures.  Each record was obtained by overlaying the shadowgraph images obtained from the high speed visualization videos of the same experiment.  The detonation wave propagated from left to the right in the diverging domain. The leading shock front is identified by the thick black line, followed by a bright, corresponding to the strong chemi-luminescence of the gas at high temperatures, which is overlaid on the shadowgraph record. 

At sufficiently high pressures (Fig. \ref{fig:AL1}(a) and (b)), the detonation front is textured with a large number of small sized cells. Traces of the transverse waves starting from the triple points on the wave front, and extending back across the reaction zone are also observed. As the detonation progresses in the enlarging section, it acquires the expected curvature due to the geometrical divergence of the gases behind the front.  
With decreasing pressure, such as Fig. \ref{fig:AL1}(c) and (d), the cellular structure of the front is enlarged, partly due to the lower reactivity of the gas at lower pressures (see below) and partly due to the stronger propensity for the detonation wave to be attenuated by the diverging geometry.  A somewhat thicker reaction zone is also observed.  The cellular structure is characterized by transverse detonations consuming the gas behind the incident shock portions of the front.  This can be clearly seen in Fig. \ref{fig:AL1}(d), where most of the gas accumulating as non-reacted behind the weaker portion of the shock near the bottom wall reacts very fast behind a transverse wave, which takes the form of a transverse detonation wave with fine cellular structure typical of near limit detonations reported by  \citet{Gamezoetal2000}.  Very few non-reacted pockets are observed, with negligible volumes, as shown in Fig. \ref{fig:AL1}(d).  When the initial pressure was lowered below a critical value, the detonation was extinguished in the diverging channel, with a quasi-1D reaction zone decoupling from the lead shock, as shown in Fig. \ref{fig:AL1}(e).  

\begin{figure}
  \centerline{\includegraphics[width=0.6\textwidth]{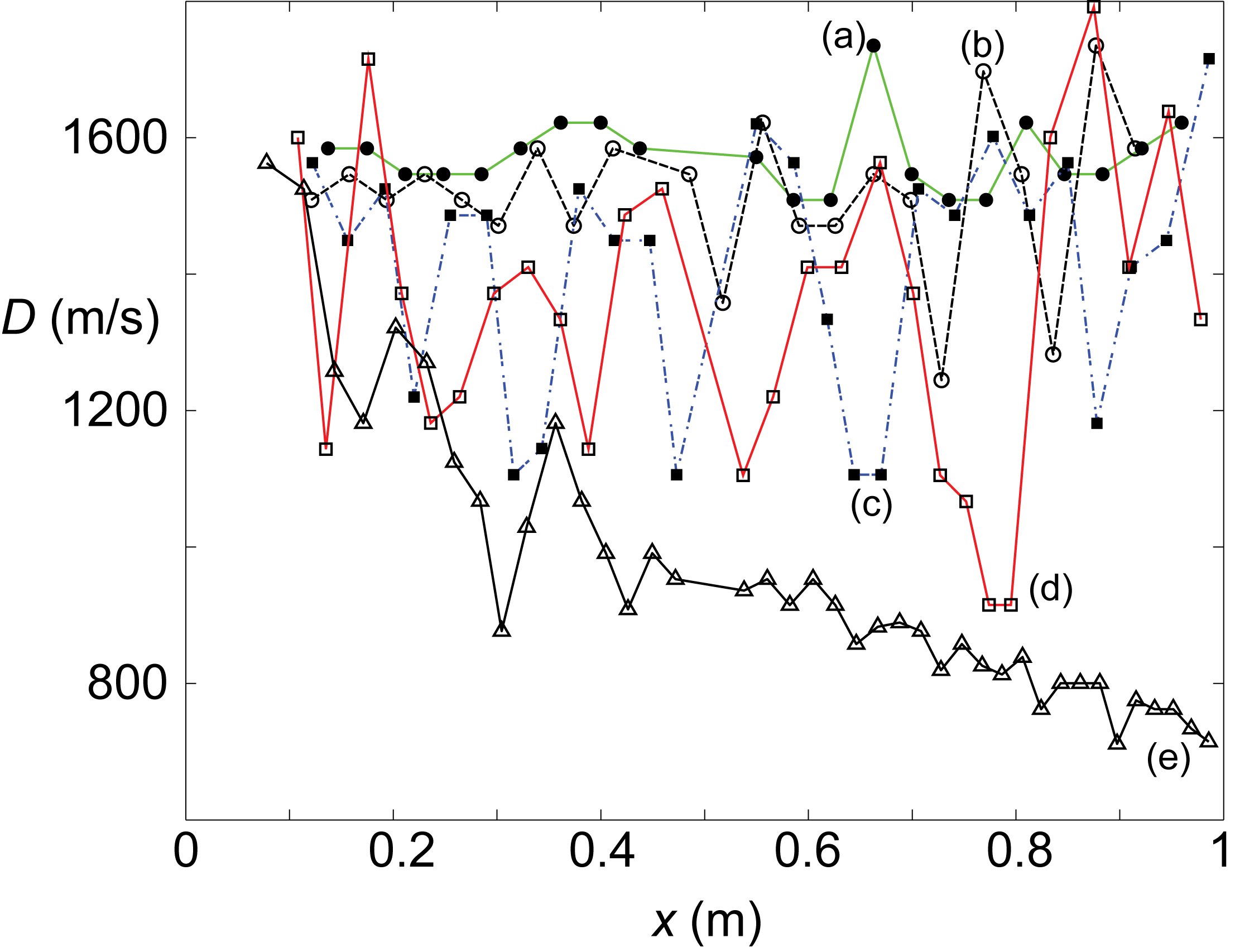}}
  \caption{Speed of detonation front recorded along the top wall in 2C$_2$H$_2$+5O$_2$+21Ar along the long ramp at initial pressures of (a) 12.1 kPa, (b) 9 kPa, (c) 6.2 kPa, (d) 4.9 kPa and (e) 4.2 kPa, corresponding to the experiments shown in Fig. \ref{fig:AL1}.}
\label{fig:AL1speeds}
\end{figure}

The speed of the leading front recorded along the top wall from the sequential frames is shown in Fig. \ref{fig:AL1speeds} for the experiments shown in Fig. \ref{fig:AL1}.  At the higher pressures, the framing rate does not resolve the cellular structure of the front, and an essentially time-averaged speed is recorded. The experiments indicate that this speed remains constant for the entire detonation travel in the diverging channel.  This confirms the concept that an exponential horn can maintain a steady travelling detonation.  

At the lower pressures of 6.2 and 4.9 kPa, the large cellular structures shown in Fig. \ref{fig:AL1}(c) and (d) are now better resolved.  The cellular dynamics give rise to speed fluctuations of approximately 500 m/s, or 30\% of the average propagation speed.  In spite of these very large fluctuations typical of cellular dynamics, the average speed is approximately constant.  Finally, at an initial pressure of 4.2 kPa, the lead shock is seen to continuously decay.  A detonation wave can no longer remain self-sustained in the diverging section - see also Fig. \ref{fig:AL1}(e) showing the absence of cellular structures.  

\begin{figure}
  \centerline{\includegraphics[width=0.7\textwidth]{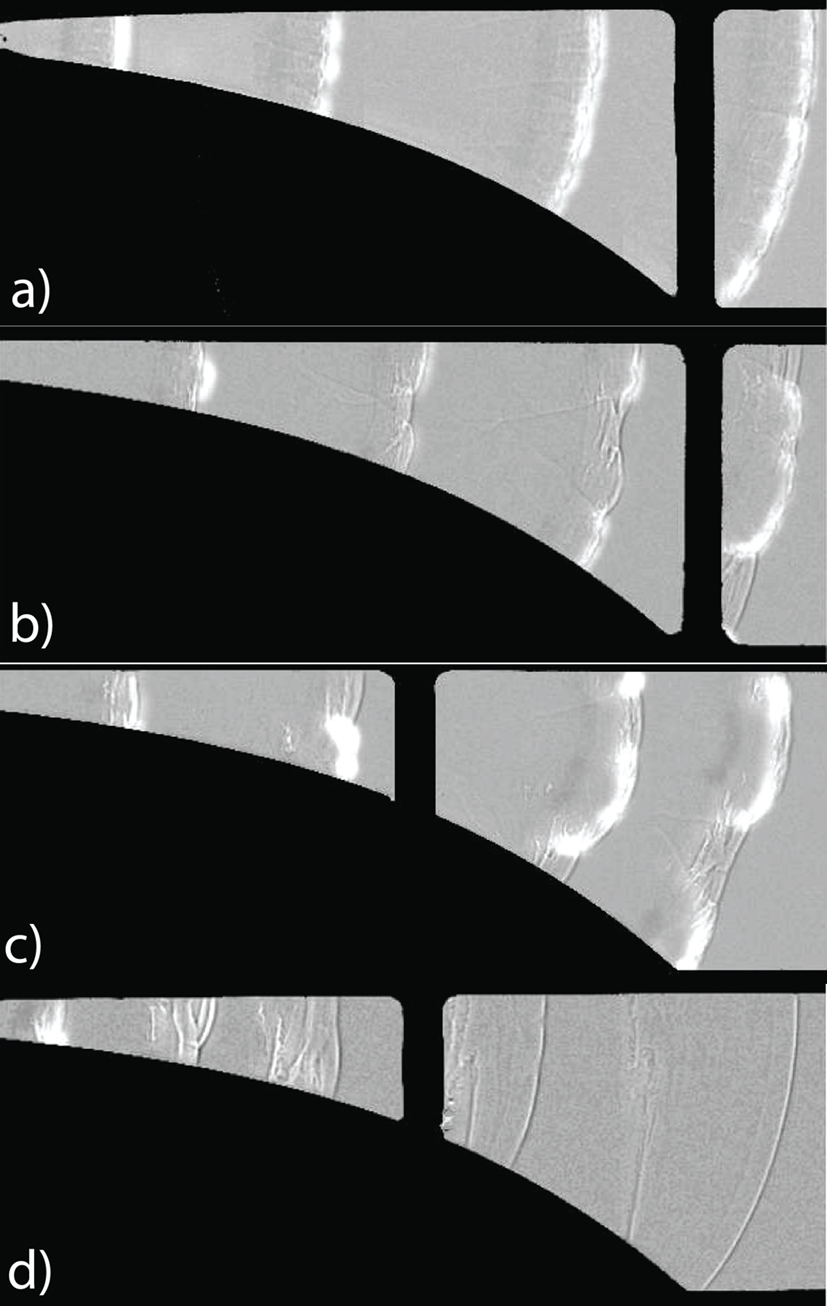}}
  \caption{Detonation front evolution in 2C$_2$H$_2$+5O$_2$+21Ar along the short ramp at initial pressures of (a) 12.1 kPa, (b) 8.3 kPa, (c) 6.2 kPa,  and (d) 4.9 kPa.}
\label{fig:AS1}
\end{figure}

\begin{figure}
  \centerline{\includegraphics[width=0.6\textwidth]{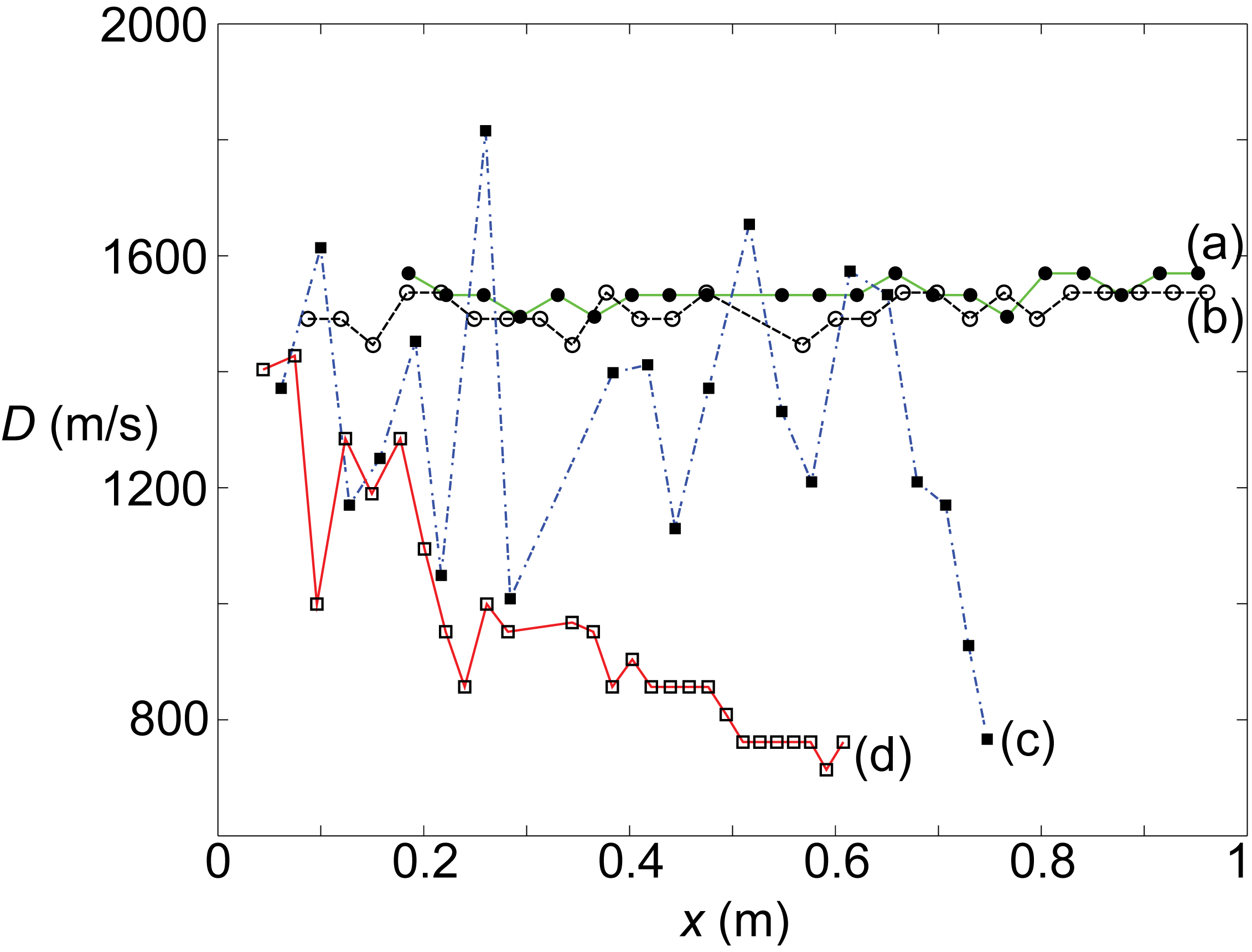}}
  \caption{Speed of detonation front recorded along the top wall in 2C$_2$H$_2$+5O$_2$+21Ar along the short ramp at initial pressures of (a) 12.1 kPa, (b) 8.3 kPa, (c) 6.2 kPa and (d) 4.9 kPa corresponding to the experiments shown in Fig. \ref{fig:AS1}.}
\label{fig:AS1speeds}
\end{figure}

The experiments performed with the shorter ramp with double the amount of divergence showed very similar flow fields. Fig. \ref{fig:AS1} shows the evolution of the front of the detonation, while Fig. \ref{fig:AS1speeds} shows the recorded speeds along the top wall for four typical experiments showing the range of phenomena observed.  While at the higher pressures, the detonation front has a uniform curvature and small cellular structures, the cellular structure enlarges at lower pressures.  The records indicate that the cellular structure corresponds to reactive transverse waves, as documented in detail by \citet{Gamezoetal2000}.  The speeds record of Fig. \ref{fig:AS1speeds} shows that for the shorter ramp, a steady detonation can also be established.  Once the cellular structure can be resolved by the framing rate used (Figs. \ref{fig:AS1}(c) and \ref{fig:AS1speeds}(c),), the front has variations of approximately 500 m/s, or 30\% of the average propagation speed.  When the pressure is sufficiently low, the detonation wave cannot propagate in the diverging section.  The front consists of a decoupled shock-flame structure (Fig. \ref{fig:AS1}(d)).  The lead shock decays to low speeds of approximately 800 m/s by the end of the channel (Fig. \ref{fig:AS1speeds}(d)).

\begin{figure}
 \centerline{\includegraphics[width=0.6\textwidth]{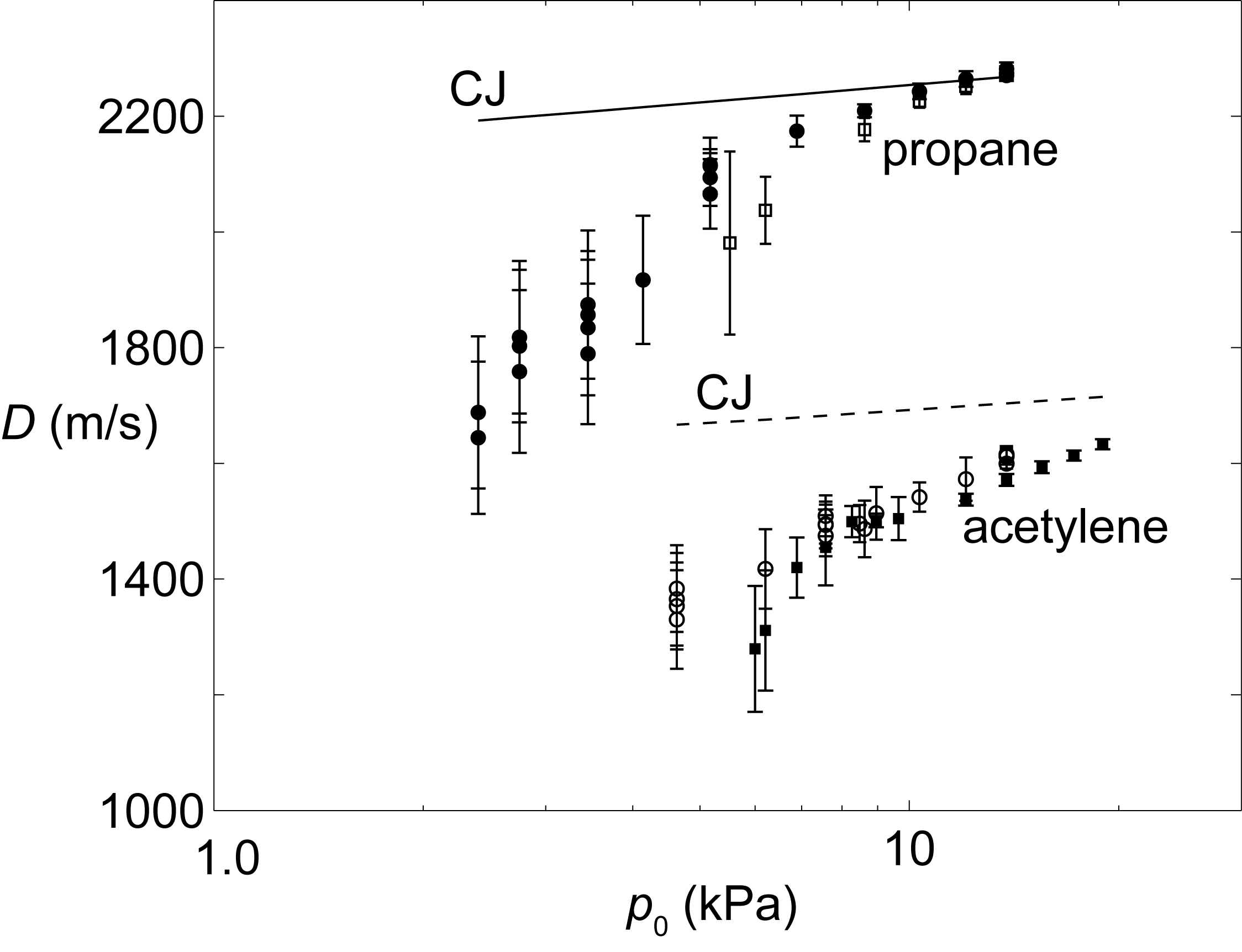}}
  \caption{Mean detonation wave speeds along the top wall for all initial pressures permitting the sustenance of detonations.}
\label{fig:D_Vs_P_0}
\end{figure}

A summary of average speeds measured in all the experiments in 2C$_2$H$_2$+5O$_2$+21Ar on both ramps are shown in Fig. \ref{fig:D_Vs_P_0} as a function of the initial pressure in the mixture.  The error bars represent the standard deviation of the measurements for each experiment.  As expected, the error bar grows with decreasing pressure, since the cellular dynamics are better resolved.  Repeat experiments are also shown.  The spread between repeat experiments never exceeded 4\%, which is significantly lower than the standard deviation due to the cellular dynamics fluctuations.  

Also shown in the Fig. \ref{fig:D_Vs_P_0} is the ideal Chapman-Jouguet (CJ) detonation speed, which was obtained from chemical equilibrium calculations using the NASA CEA code \citep{CEA}. The figure shows that both the CJ speed and the experimentally determined speed drop for both large and small ramp experiments when the initial pressure is lowered. The CJ speed drops because of the enhanced dissociation in the product species at low pressures. Nevertheless, the mean shock speed recorded in the experiments also deviates from the CJ speed with a decrease in the initial pressure. This can be interpreted in terms of the gas sensitivity varying with the initial pressure. Mixtures with lower initial pressures are the ones with slower reaction rates and chemical energy release rate - see Fig. \ref{fig:kinetictimes}. Therefore, the expansion cooling experienced by the shocked gas particles due to the lateral area divergence is able to decrease the shock strength and speed. Since the CJ speed represents the velocity of a steady detonation wave, the lower initial pressures result in more deviation from the CJ speed.

\subsection{Results in C$_3$H$_8$+5O$_2$}
\begin{figure}
  \centerline{\includegraphics[width=1\textwidth]{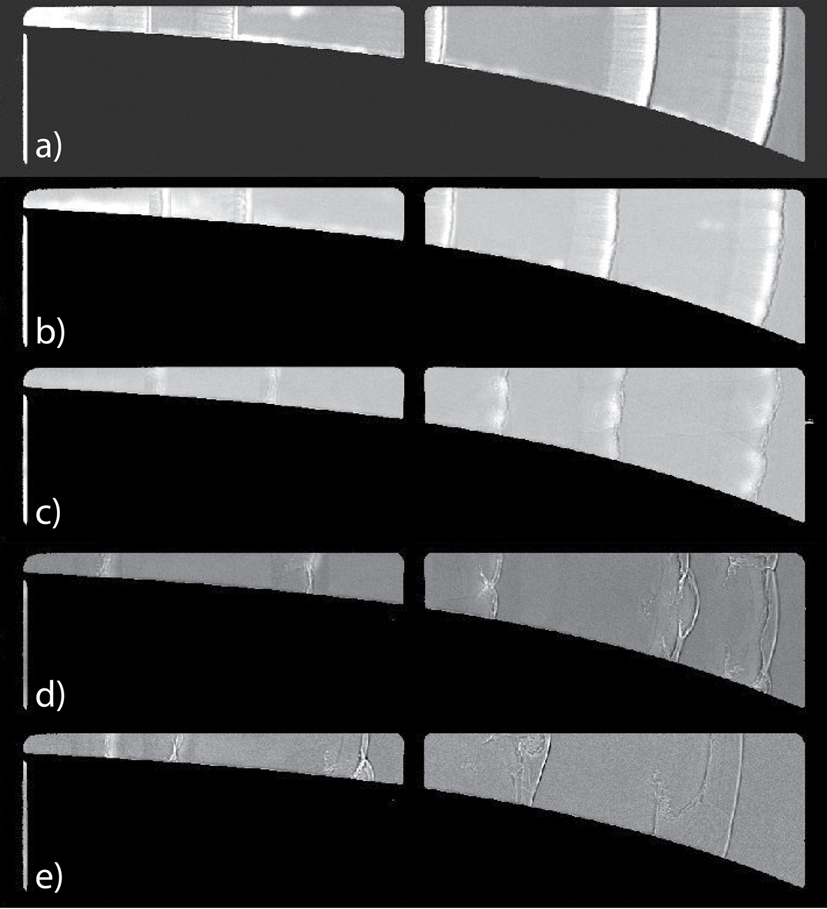}}
  \caption{Detonation front evolution in C$_3$H$_8$+5O$_2$ along the long ramp at initial pressures of (a) 12.1 kPa, (b) 8.7 kPa, (c) 5.2 kPa, (d) 3.4 kPa and (e) 2.8 kPa.}
\label{fig:PL1}
\end{figure}

\begin{figure}
  \centerline{\includegraphics[width=0.6\textwidth]{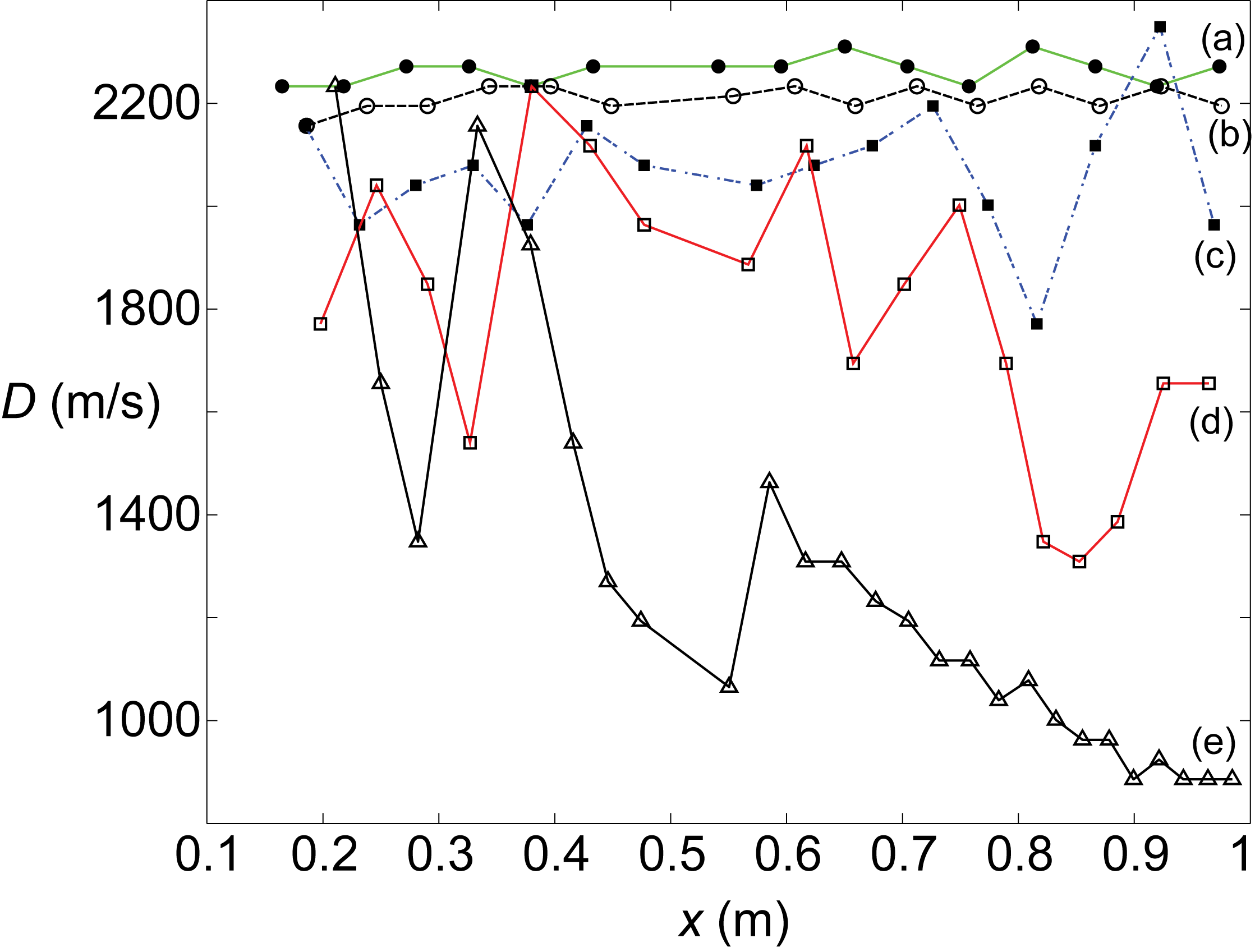}}
  \caption{Speed of detonation front recorded along the top wall in C$_3$H$_8$+5O$_2$ along the long ramp at initial pressures of (a) 12.1 kPa, (b) 8.7 kPa, (c) 5.2 kPa, (d) 3.4 kPa and (e) 2.8 kPa, corresponding to the experiments shown in Fig. \ref{fig:PL1}.}
\label{fig:PL1speeds}
\end{figure}

\begin{figure}
  \centerline{\includegraphics[width=0.6\textwidth]{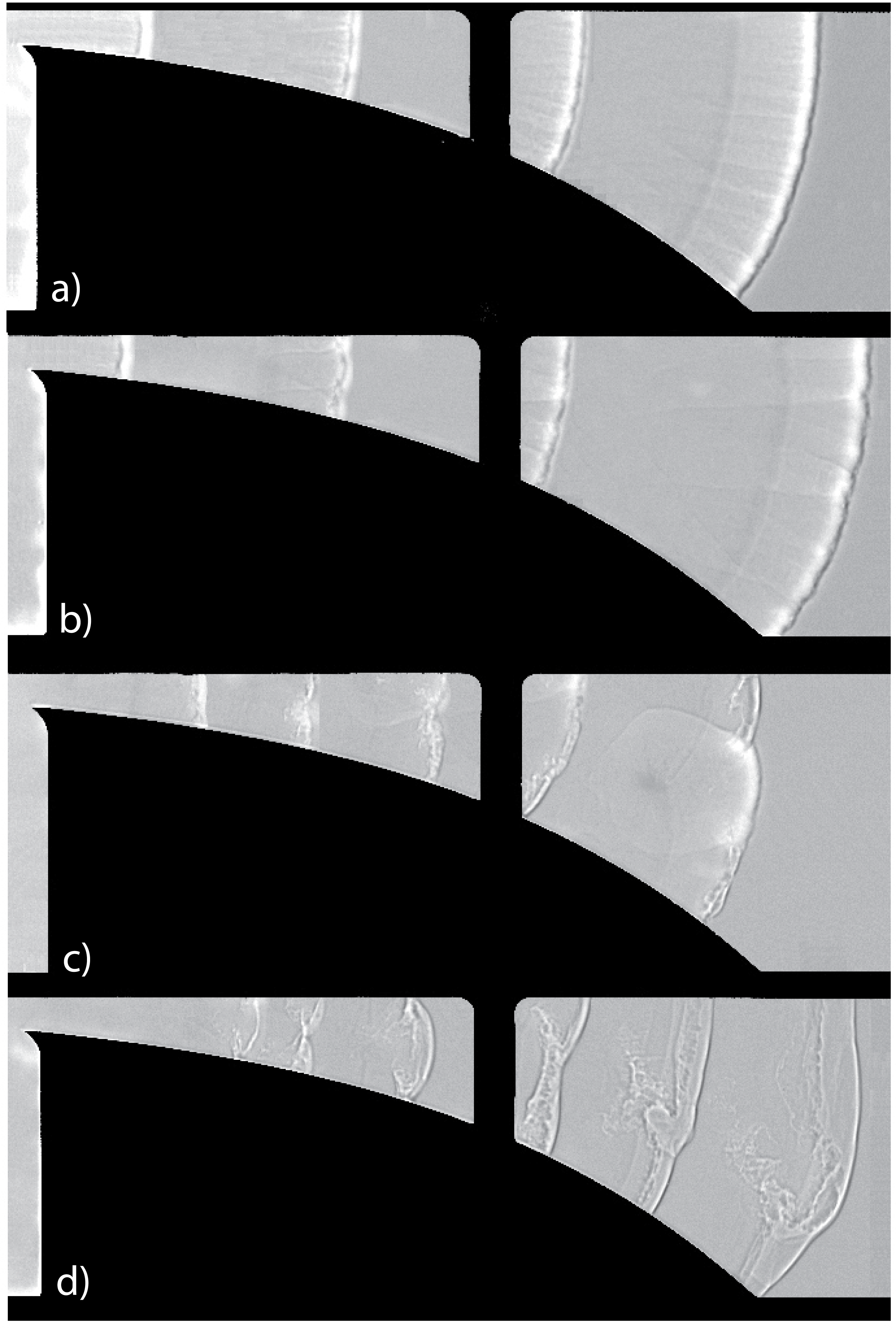}}
  \caption{Detonation front evolution in C$_3$H$_8$+5O$_2$ along the long ramp at initial pressures of (a) 12.1 kPa, (b) 8.6 kPa, (c) 5.5 kPa, and (d) 4.5 kPa.}
\label{fig:PS1}
\end{figure}

\begin{figure}
  \centerline{\includegraphics[width=0.6\textwidth]{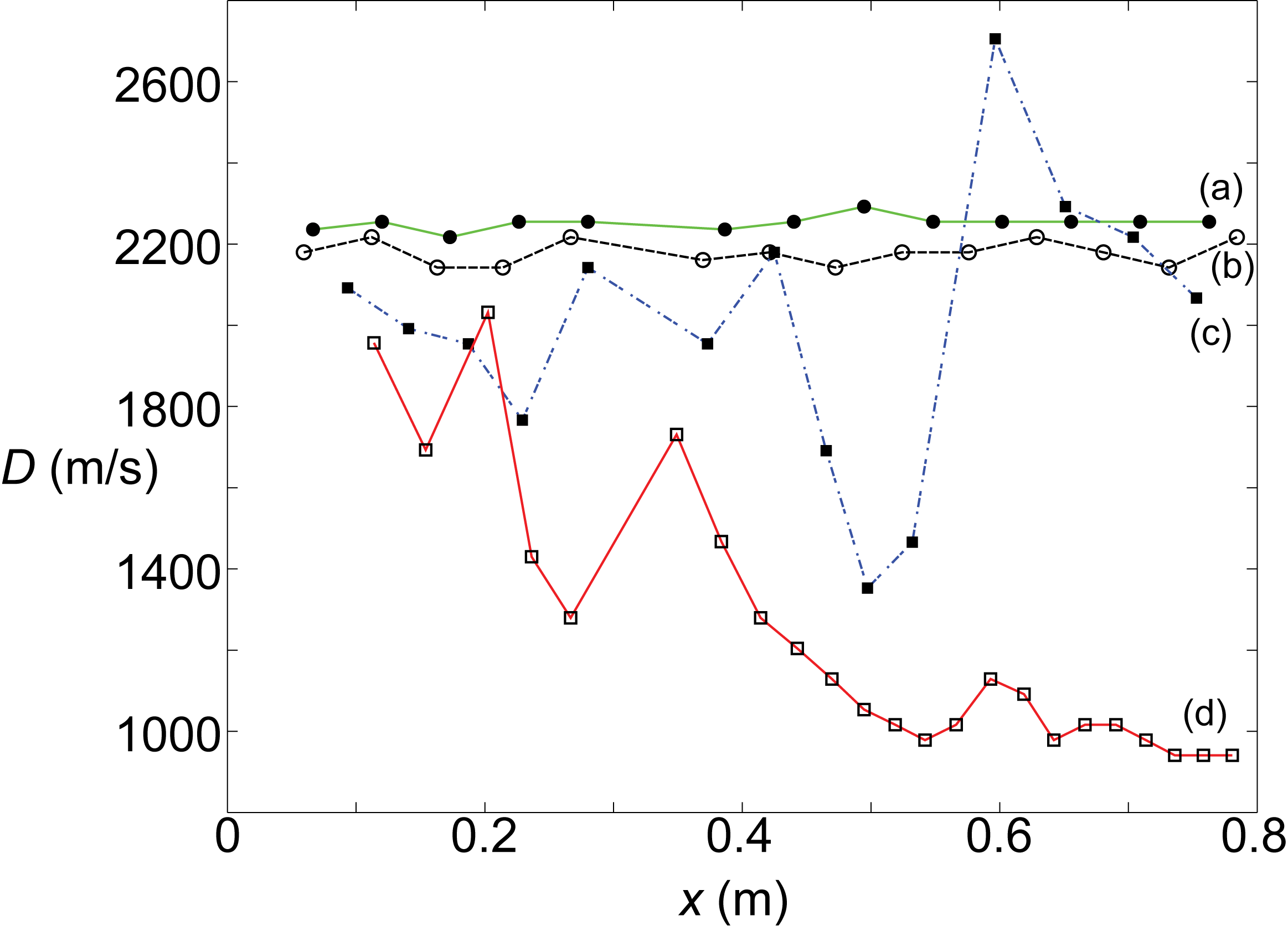}}
  \caption{Speed of detonation front recorded along the top wall in C$_3$H$_8$+5O$_2$ along the shorter ramp at initial pressures of (a) 12.1 kPa, (b) 8.6 kPa, (c) 5.5 kPa, and (d) 4.5 kPa, corresponding to the experiments shown in Fig. \ref{fig:PS1}.}
\label{fig:PS1speeds}
\end{figure}

The experiments performed in the C$_3$H$_8$+5O$_2$ mixture were qualitatively similar to the ones presented above for the acetylene mixture - although some fundamental differences were observed.  Figs. \ref{fig:PL1} and \ref{fig:PL1speeds} show the results for the long ramp, while Figs. \ref{fig:PS1} and \ref{fig:PS1speeds} show the results for the short ramp.  Firstly, for these mixture, a steady detonation was again observed in the diverging sections, confirming the appropriateness of a mean field description to model the dynamics.

\begin{figure}
  \centerline{\includegraphics[width=1\textwidth]{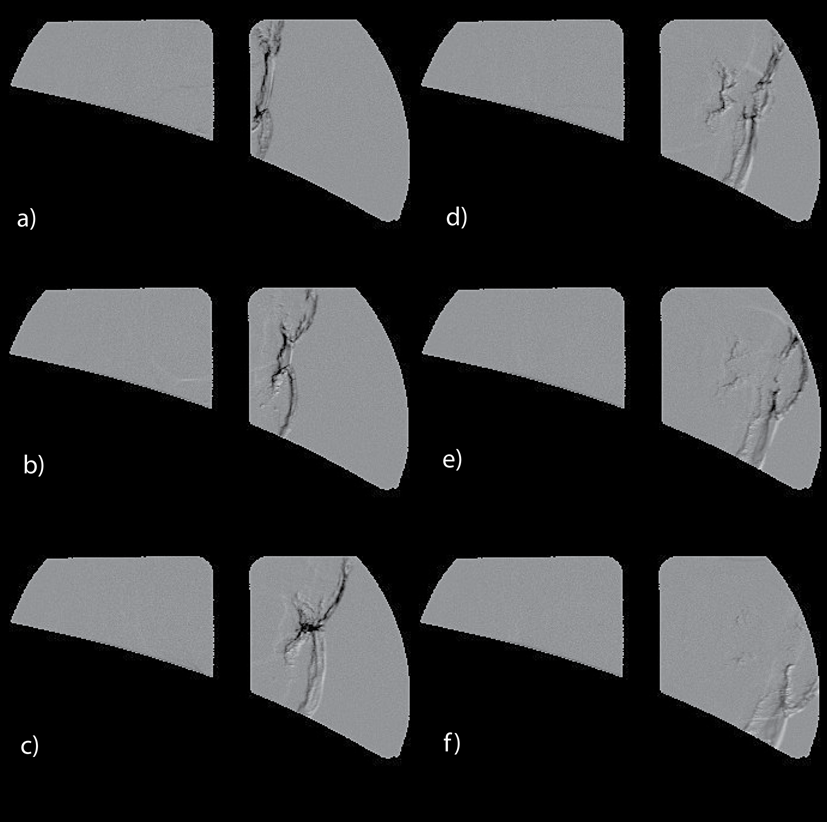}}
  \caption{Sequence of Schlieren records illustrating the front evolution in C$_3$H$_8$+5O$_2$ along the short ramp at 4.5 kPa.}
\label{fig:PS2}
\end{figure}

The most important difference with the earlier mixture was the cellular mode of propagation, as shown in Fig. \ref{fig:PL1}(d) at a sufficiently low pressure such that the cells are large enough to be well resolved.  The transverse waves for this mixture appeared as non-reactive, with tongues of unreacted pockets extending downstream and sometime pinched off as pockets.  This is typical of more unstable detonations \citep{Austin2003, Subbotin1976, Radulescuetal2007, maxwell2017influence}. Nevertheless, reactive transverse waves were also sometimes observed, as shown in Fig. \ref{fig:PS1}(c). These were however the exception.  
In order to fully address the mode of propagation of these more unstable detonations, a few Schlieren experiments were also performed for the small ramp, since the field of view of these was limited to the dimension of the mirrors (30 cm).  The set-up for the Schlieren experiments is described elsewhere \citep{Bhattacharjee2013}.  Fig. \ref{fig:PS2} shows part of the front evolution for an initial pressure of 4.8 kPa.  As can be clearly seen at this time and space resolution, the detonation wave takes on a characteristic cellular structure with non-reactive transverse waves \citep{Austin2003, Subbotin1976, Radulescuetal2007, maxwell2017influence}.  Gases accumulating behind the incident shock, here the central part of the front in Fig. \ref{fig:PS2}(a) accumulate as non-reacted layers of gas, which are then pinched off from transverse wave collisions (Fig. \ref{fig:PS2}(c)).  These pockets are then thinned out from their edges inwards through diffusive phenomena (flames).     

A summary of the average speeds measured in all the experiments in C$_3$H$_8$+5O$_2$ on both ramps are shown in Fig. \ref{fig:D_Vs_P_0} as a function of the initial pressure in the mixture, along with the ideal Chapman-Jouguet speeds calculated with the chemical equilibrium code CEA.  Similar to the acetylene experiments, the velocity deficit grows with a decreasing pressure until extinction is observed.

\subsection{The experimental $D(K)$ curves}    
The total mass divergence rate experienced by the detonation wave in the experiments of this study is due to the area divergence due to the diverging channel \textit{and} the divergence of the flow to the boundary layers on the channel walls inherent in a thin channel geometry, by the mechanism discussed in the introduction.  The effective lateral flow divergence rate thus includes both the contribution from the the diverging geometry and and the contribution of the flow divergence rate due to the boundary layers developing on the walls of the channel,$\phi_{BL}$:  

\begin{equation}
 K_{eff}=\frac{1}{A}\frac{dA}{dx}+\phi_{BL}
  \label{eq6-1}
\end{equation} 

\begin{figure}
  \centerline{\includegraphics[width=0.8\textwidth]{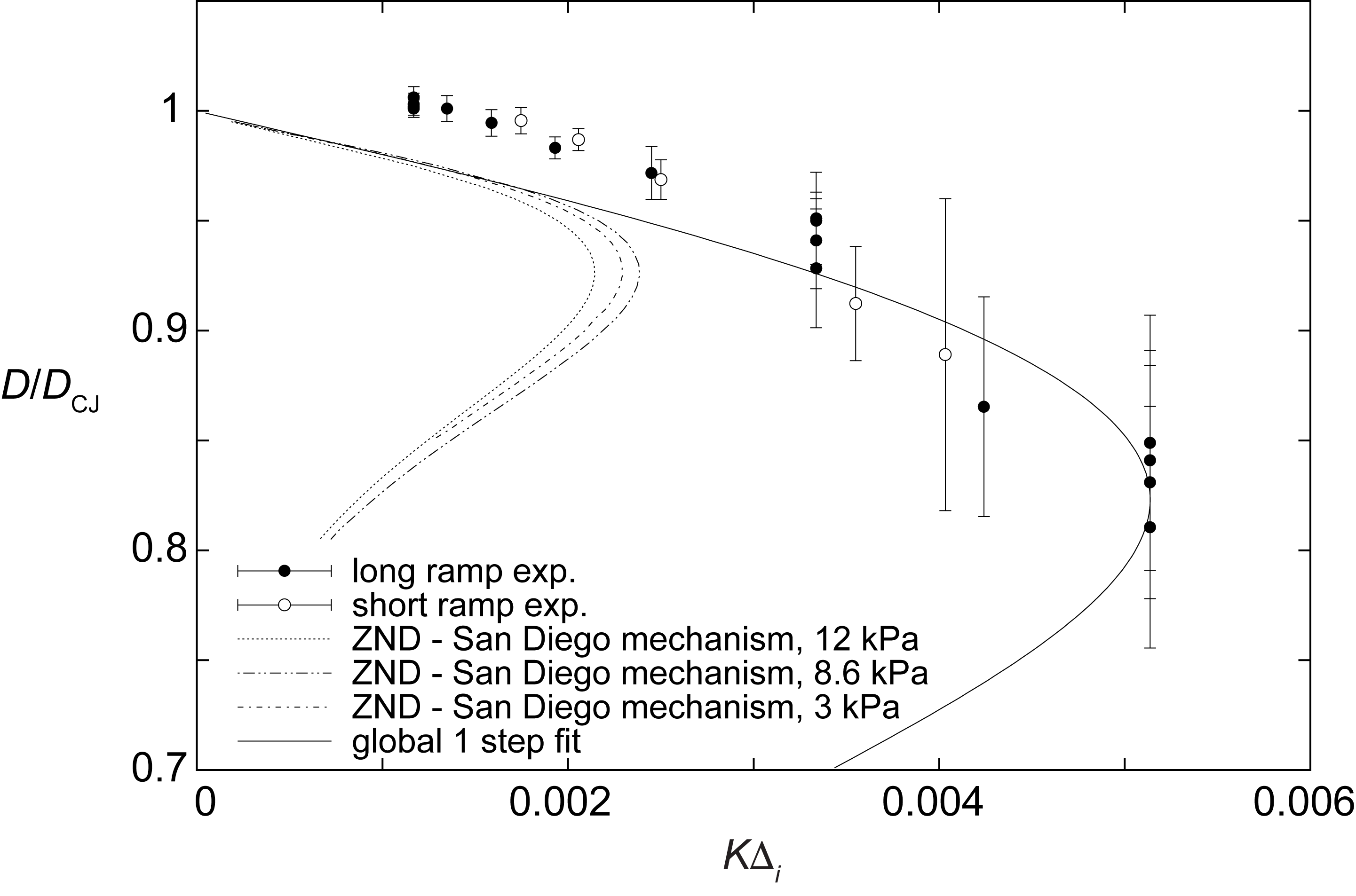}}
  \caption{The $D-K$ characteristic curves for C$_3$H$_8$+5O$_2$ obtained experimentally and predicted from the quasi-one-dimensional ZND model using the San Diego chemical mechanism \citep{SanDiego}; solid line is an empirical fit discussed in section 5.2.}
\label{fig:Comparison_Propane}
\end{figure}

\begin{figure}
  \centerline{\includegraphics[width=0.8\textwidth]{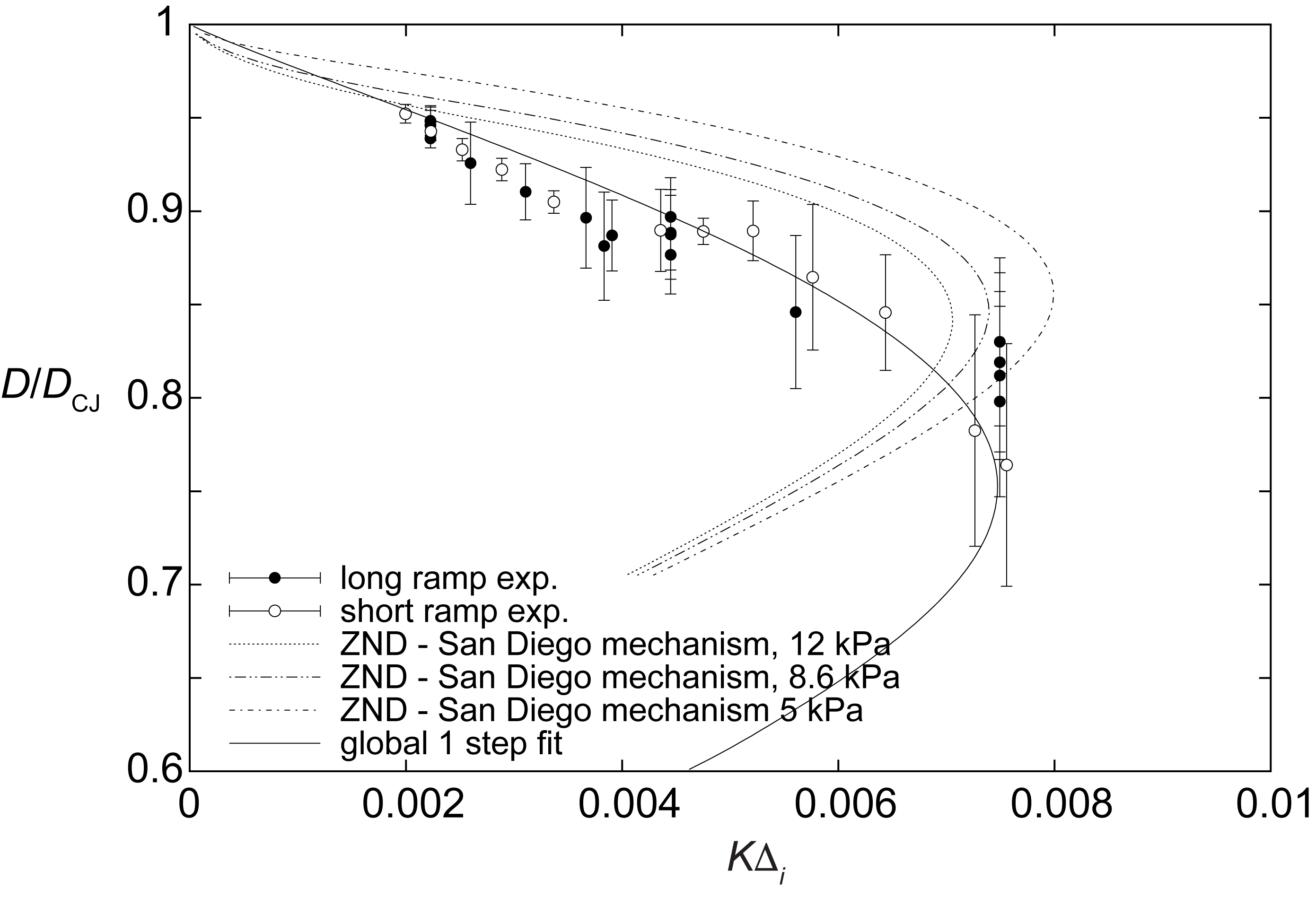}}
  \caption{The $D-K$ characteristic curves for 2C$_2$H$_2$+5O$_2$+21Ar obtained experimentally and predicted from the quasi-one-dimensional ZND model using the San Diego chemical mechanism \citep{SanDiego}; solid line is an empirical fit discussed in section 5.2.}
\label{fig:Comparison_Acetylene}
\end{figure}

\begin{figure}
  \centerline{\includegraphics[width=0.7\textwidth]{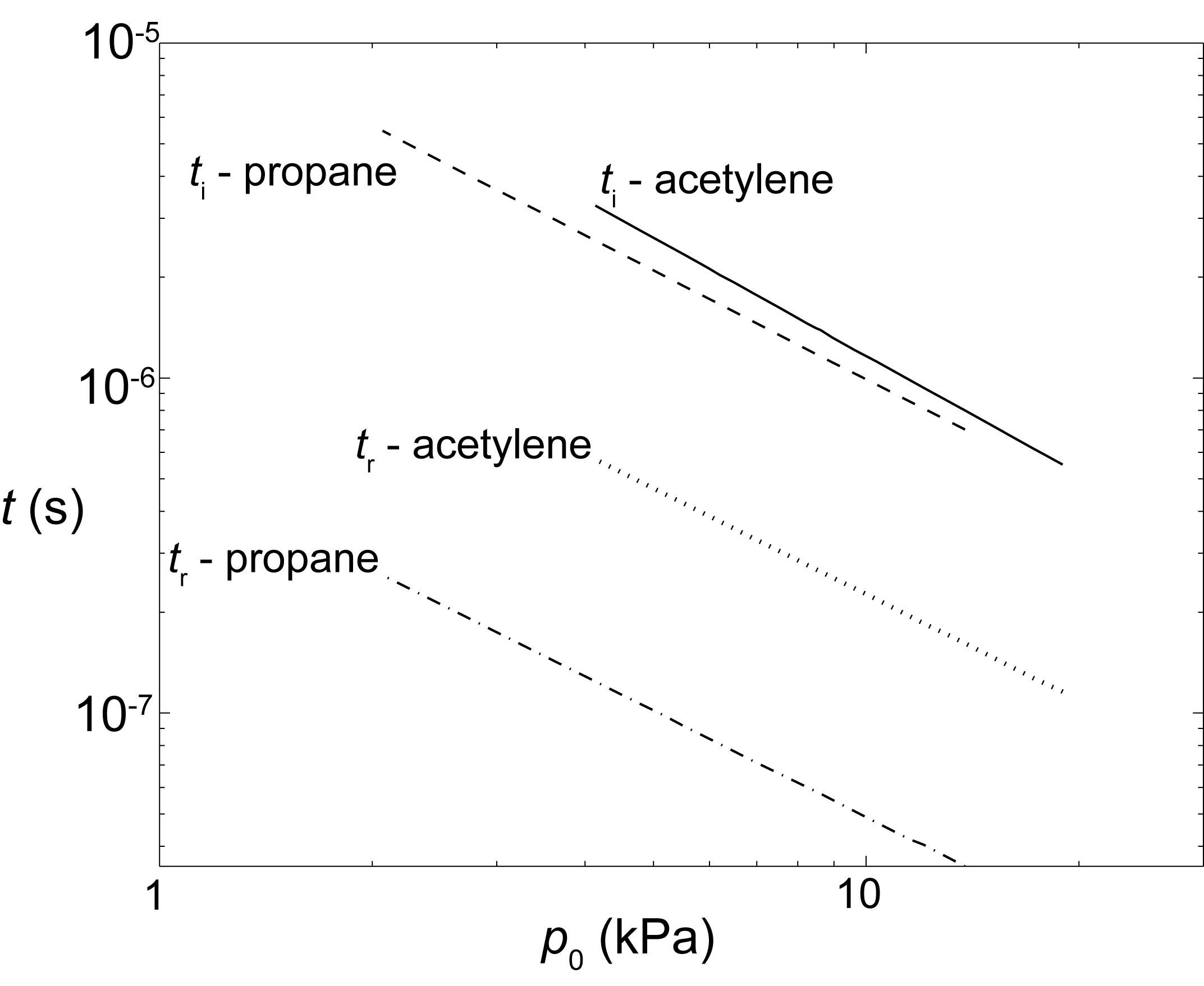}}
  \caption{Variation of the ignition delay time $t_{i,N}$ and reaction time $t_{r,N}$ for the two mixtures with initial pressure calculated at the post shock conditions of the steady ZND structure in the absence of losses.}
\label{fig:kinetictimes}
\end{figure}

Instead of modelling the source term $\phi_{BL}$, which cannot be done without an empirical constant \citep{Chaoetal2009, Camargoetal2010, Gaoetal2016}, this loss rate can be directly evaluated from the experiments conducted on the two ramps of different expansion ratios, but constant channel dimension (19 mm), leading to a constant $\phi_{BL}$. The contribution of the boundary layers losses $\phi_{BL}$ has been obtained experimentally by comparing the experiments performed on the two ramps, for the same mixture, and same velocity deficit and calibrating the effective rate of mass divergence $K_{eff}$ to obtain a unique relation between velocity deficit and loss rate, as expected from theoretical considerations.  Figures \ref{fig:Comparison_Propane} and \ref{fig:Comparison_Acetylene} show the data obtained after such a reduction of the data shown in Fig. \ref{fig:D_Vs_P_0}.  The induction length used to normalize the lateral divergence parameter for each mixture was calculated as the product of the flow speed at the Von Neumann post shock state in the detonation frame of reference and the ignition delay time calculated for constant volume combustion at the Von Neumann state using the San Diego mechanism, which is shown in Fig. \ref{fig:kinetictimes}. It was found that a unique value of $\phi_{BL}=5.5 m^{-1}$ permitted to collapse all data for the two ramps in both mixtures tested. Note that this value is comparable with the divergence provided by the exponential horns themselves.  Note also that the reaction time shown in Fig. \ref{fig:kinetictimes}, calculated as maximum inverse thermicity in the calculations, varies also proportionally to the induction time in each mixture, such that the normalization by this time scale would have generated the same result.   

\section{Discussion}
\subsection{Comparison of $D(K)$ curves obtained experimentally with the ZND model for real gases}

Figures \ref{fig:Comparison_Propane} and \ref{fig:Comparison_Acetylene} also show the comparison of the $D(K)$ curves obtained experimentally with the curves obtained from the ZND model with area divergence, described in Section 2.  For both mixtures, the solution was obtained using the San Diego chemical kinetic mechanism.  The calculations were performed at three initial pressures spanning the experimental range.  For each pressure, the lateral divergence was varied.  Normalizing the curves by the induction lengths permitted to collapse the data very well.  

For the 2C$_2$H$_2$+5O$_2$+21Ar mixture, the experiments show that the detonation speeds are reliably lower than the predictions.  These cannot be accounted for by limitations of the exponential horn in maintaining a constant divergence rate, as this would have the opposite effect - see section 3. The limiting value of maximum lateral mass divergence predicted by the steady model captures the experimental limit quite well.   

For the more unstable C$_3$H$_8$+5O$_2$ detonations, the agreement between experiment and the steady model obtained from the real thermo-chemical data is quite poor.  Overall, the experiments show that the detonation speeds are always larger than predicted at a given global divergence rate.    Moreover, the experiments also show that the limiting value of divergence is larger by approximately a factor of 2-3 than predicted, while the maximum velocity deficit is also 3 times larger in the experiments. While such discrepancies may be argued to be a signature of faulty kinetics, we do not think this is the reason.  Incorrect predictions of kinetics time scales would only shift the prediction curve left or right, i.e., re-scale the abscissa, but would not affect the velocity deficits, which are a very strong function of the sensitivity of the kinetics, i.e., via the global activation energy.  Instead, we believe that the large discrepancy between experiments and predictions is due to the burning mechanism in the highly-unstable detonations, which is not captured in the model.  As shown in Fig. \ref{fig:PS2}, layers of shocked, non-reacted gas reacts by diffusive processes (flames), consistent with the results of \cite{maxwell2017influence} and \cite{Radulescuetal2007}.  This phenomenon is to reduce the thermal sensitivity of the global thermicity, as it is shown empirically below.  

Another interesting observation of the experiments is the extrapolation of the detonation speed measured experimentally to zero divergence in Figs. \ref{fig:Comparison_Acetylene} and \ref{fig:Comparison_Propane}.  For the acetylene mixture, the experimental data appear to extrapolate well to the predicted CJ speed.  For the more unstable detonations in propane, the experimental data extrapolate to speeds in excess of the CJ prediction by a few percent.  This result is also compatible with previous measurements, which typically report propagation speeds, when extrapolated to zero losses, larger than the CJ predictions by a few percent - see \citet{Fickett&Davis1979} for discussion and earlier references. 

The results obtained for the C$_3$H$_8$+5O$_2$ detonations unambiguously show that cellular detonations are more detonable than predicted by the steady ZND model.  This experimental fact appears at odds with current numerical simulations of inviscid cellular detonations, which have showed the opposite trend \citep{RadulescuetalICDERS2007, han2017role, han2017effect, reynaud2017computational, Mazaheri2015}.  While the authors found pockets of non-reacted gas as the reason for the delay of the global exothermicity, our current experiments show that these pockets burn via surface flames very fast - see also \citep{maxwell2017influence} for a quantitative analysis.  

\subsection{Effective reaction models for cellular detonations.}
From a more practical perspective, the experimental technique developed in the present study of determining the detonation speed relation on the area divergence can be used to tune macroscopic models for the detonation dynamics, in a somewhat similar fashion to condensed phase reaction modelling \citep{Bdzil&Stewart2007}.  One of the simplest models that can be postulated is the one-step model discussed earlier in Sections 2 and 3, although now its parameters are taken to represent the global behaviour of detonations at macro-scales.   It is instructive to comment on the magnitude of the fitting constants, which in this case are the effective activation energy, the pre-exponential factor in the rate equation, the isentropic exponent and the chemical heat release. 

The isentropic exponent $\gamma$ was chosen to match the value obtained using realistic thermo-chemical data behind the lead shock propagating at the CJ speed, in order to correctly recover the gas compressibility in the reaction zone of detonations.  The heat release $Q$ was chosen such that the detonation Mach number that can be calculated using the perfect gas model \citep{Lee2008} matched the calculation result using the exact thermo-chemical data for that particular mixture. These are listed in Table \ref{tab:table_parameters_fit}.

The curve fitting for the rate parameters focused on matching the turning point of the $D(K)$ curves obtained experimentally.  The velocity deficit is a strong function of the activation energy, while the lateral divergence at the turning point can be re-scaled by tuning of the pre-exponential factor in the one-step rate equation \citep{Yao1995, Kleinetal1994, He1994}.  Figures\ref{fig:Comparison_Propane} and \ref{fig:Comparison_Acetylene} show the resulting fits, resulting in the parameters listed in Table \ref{tab:table_parameters_fit}.  A very good agreement can be obtained with experiment empirically for both mixtures.  The parameter $G$ listed in Table \ref{tab:table_parameters_fit} represents the scale factor for the non-dimensional pre-exponential factor, when the half reaction length is used in the one step model and the induction zone length is used to reduce the experimental data.  

The calibration also highlights what the effective activation energies are in these fits, as compared with those derived from the underlying full chemistry.  The real activation energies were obtained in the usual way. The activation energy was extracted from the assumed dependence of the ignition delay $t_{i} \sim \exp \left(\frac{E_a}{RT}\right)$ obtained at the Von Neumann state of the CJ detonation, yielding:

\begin{equation}
 \frac{E_a}{RT_N}=\frac{1}{T_N} \left(\frac{d \ln t_i}{d (1/T)}\right)_{T=T_N}
  \label{eq:activationenergy}
\end{equation}
\noindent The ignition delay calculations were performed at constant volume using Cantera and the San Diego kinetic mechanism.  The derivative was estimated numerically by bracketing the Von Neumann state by 100 K. 

The activation energies tabulated in Table \ref{tab:table_parameters_fit} show that the global decomposition kinetics in an average description of cellular detonations have activation energies lower than the underlying chemical ones.  For the acetylene mixture, the reduction in effective activation energy is 14\%, while for propane, the reduction in the activation energy is by 54\%!  The significantly lower effective activation energy for the much more unstable propane mixture is not surprising, as it was already noted that the enhancement of burning mechanism in turbulent detonations by turbulent mixing suppresses much of the thermal character of the ignition mechanism, where the reaction zone length does not grow with velocity deficit, as anticipated from thermal ignition considerations \citep{Radulescuetal2007, maxwell2017influence}.  Instead, the rapid burn-out of non-reacted pockets by diffusive processes explains why the reaction zones remain substantially shorter. 

The effective reaction rates for the dynamics of cellular detonations extracted from these experiments can naturally find use in problems in which quasi-steady and stationary cellular detonation waves appear, for example in problems of detonations weakly confined by an inert layer \citep{reynaud2017computational} as for rotating detonation engines.  Future work should be devoted to this validation.  Interestingly, however, the effective reaction rate derived here from experiment has been suggested to be relevant for predicting highly unsteady one-dimensional problems of detonation dynamics.  The first author reports in a preliminary study that the initiation energy reported in experiments by others for blast initiation can be reproduced with a one-dimensional calculation using the fit of this study within a factor of 2  \citep{radulescu2017usefulness}.  This suggests that the effective reaction rate measured by our technique may capture the effective kinetics of cellular structure dynamics.  Clearly, much more work is required to test the relevancy of a one-step fit derived from the experiments as global kinetics for detonation dynamics.  

Nevertheless, the exponential horn experiments can in the future serve as an experimental benchmark for models for the detonation structure with various complexities. In our opinion, the fact that the detonation waves can be maintained in quasi-steady state with a global one-dimensional average structure can serve as a unique benchmark for models, both steady and unsteady, ranging from engineering type models using a single step chemistry to direct numerical simulations with complex thermo-chemical descriptions.  Indeed, the gas phase detonation community lacks a well defined experimental benchmark for validating models. Providing such a framework was the motivation of our study.

\begin{table}
 \begin{center}
 \caption{Thermo-chemical parameters for a global description of cellular detonations fitted from the exponential ramp experiments.\label{tab:table_parameters_fit}}
  \begin{tabular}{lccccc}
  \hline
     Mixture  &  $\gamma$  & $\frac{Q}{RT_0}$ & $G$ & $\frac{E_a}{RT_0}$ & $\frac{E_a}{RT_0}$ (full chem.)  \\
    \hline
       2C$_2$H$_2$+5O$_2$+21Ar   & 1.42 & 18.3 & 1.2  & 24 & 27.8\\
       \hline
       C$_3$H$_8$+5O$_2$ & 1.14 & 99 & 2.05 & 26 & 56 \\
       \hline
  \end{tabular}
 \end{center}
\end{table}
\section{Conclusion}
The present work showed that an exponential horn technique can be used to generate detonations with a constant mean lateral mass divergence.  Multi-dimensional simulations of weakly unstable detonations and experiments have demonstrated that cellular detonations had a constant mean propagation speed in the diverging section.  This confirms that a macroscopic mean field description is worthwhile for both weakly unstable and strongly unstable cellular gaseous detonations.  

The experimentally obtained detonation speed dependence on lateral mass divergence was compared with predictions using the steady ZND model with lateral divergence using real chemistry.  For the less unstable detonations in 2C$_2$H$_2$+5O$_2$+21Ar, the cellular detonations displayed somewhat larger velocity deficits than predicted, although the limiting value of maximum divergence was found in excellent agreement with the model.  For the more unstable detonations in C$_3$H$_8$+5O$_2$, the cellular detonations were found to propagate at higher speeds and to much larger divergence rates than predicted by the steady model.  The enhancement of detonability in highly unstable detonations is attributed to the importance of diffusive processes in the burn-out of the the non-reacted pockets during the propagation, as observed from the reaction zone visualization.  Furthermore, the empirical tuning of a global one-step model to describe the experimental trends revealed that the effective activation energy was lower by 54\% than the value derived from the real kinetics of the fuel decomposition.  This supports the view that diffusive processes in highly unstable detonations may be responsible for reducing the thermal ignition character of the gases processed by the detonation front \citep{Radulescuetal2007}.

We wish to acknowledge the NSERC Discovery Grant to M.I.R. and partial support from the NSERC Hydrogen Canada (H2CAN) Strategic Research Network for supporting B.B.. We would like to acknowledge Professor S. Falle of the University of Leeds for providing the numerical platform used in the study. We would also like to thank Karl German, Terrence R. Phenix, Mohammed Saif Al Islam and Qiang Xiao for help in conducting the experiments.

\section*{Appendix: resolution study}
\begin{figure}
  \centerline{\includegraphics[width=1\textwidth]{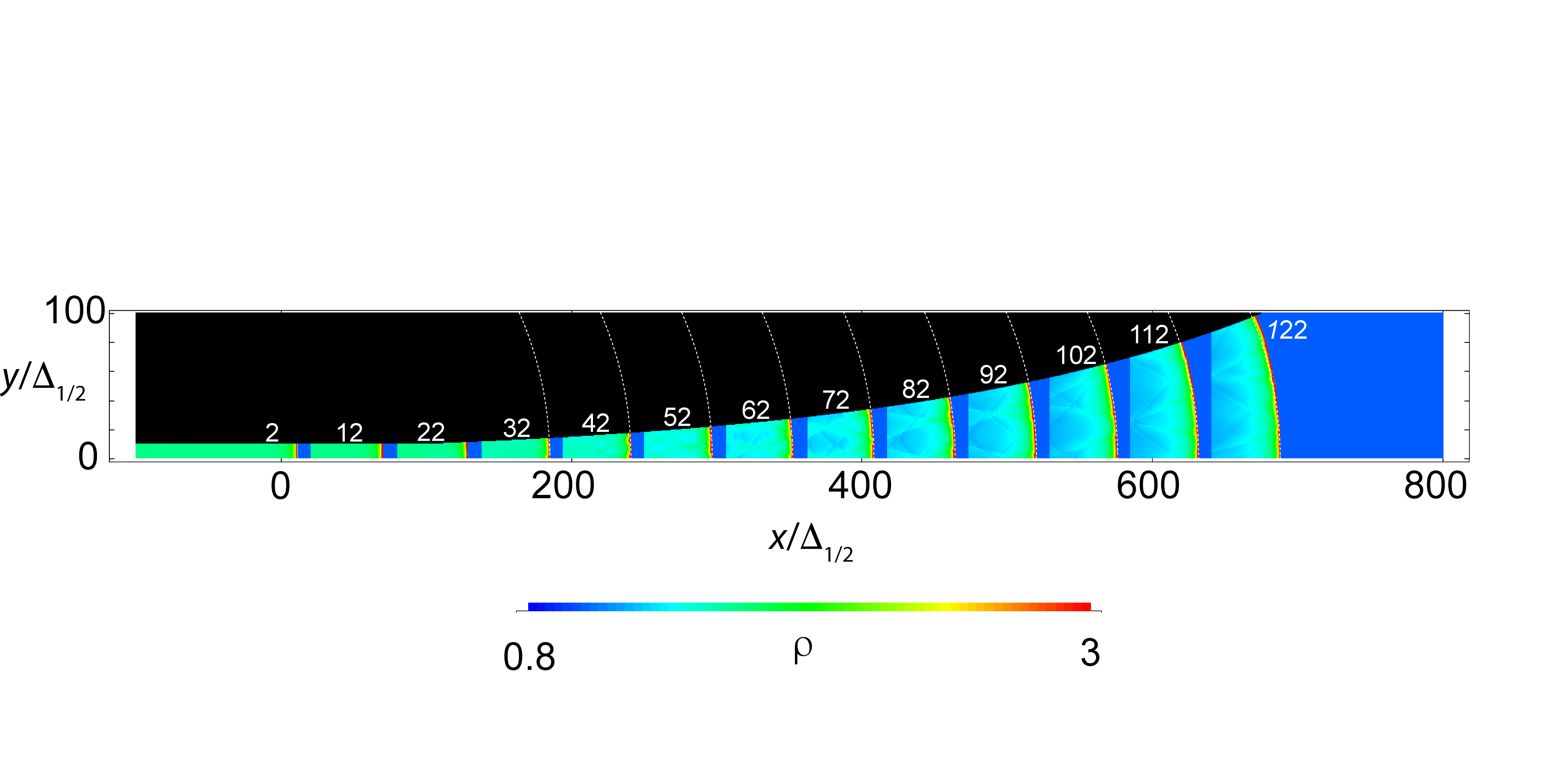}}
  \caption{Evolution of a curved detonation in an exponential horn with $\bar{K}$=0.004 with a minimum resolution of 1/32$\Delta_{1/2}$; white lines denote arcs of circle with the expected curvature; numerals indicate the non-dimensional times.}
\label{fig:K004_32pts}
\end{figure}

In order to test the effect of the numerical resolution on the results presented in section 3, we have performed a systematic resolution study by varying the minimum grid spacing over more than two orders of magnitude. Note that this minimum grid spacing is enforced over the entire reaction zone structure. The particular geometry investigated is illustrated in Fig. \ref{fig:K004_32pts}.  The domain extends from -100$\Delta_{1/2}$ to 800$\Delta_{1/2}$ in the $x$-direction and 0 to 100$\Delta_{1/2}$ in the $y$ direction.  A channel 10$\Delta_{1/2}$ - wide begins expanding at $x$ = 100$\Delta_{1/2}$ with a divergence rate of $\bar{K}$=0.004. The calculations were initialized with the ZND solution, with the lead shock positioned at $x$ = 0.  The minimum grid spacing $\delta$ was changed from 2$\Delta_{1/2}$ to 1/64$\Delta_{1/2}$ in increments of 2.

\begin{figure}
  \centerline{\includegraphics[width=1\textwidth]{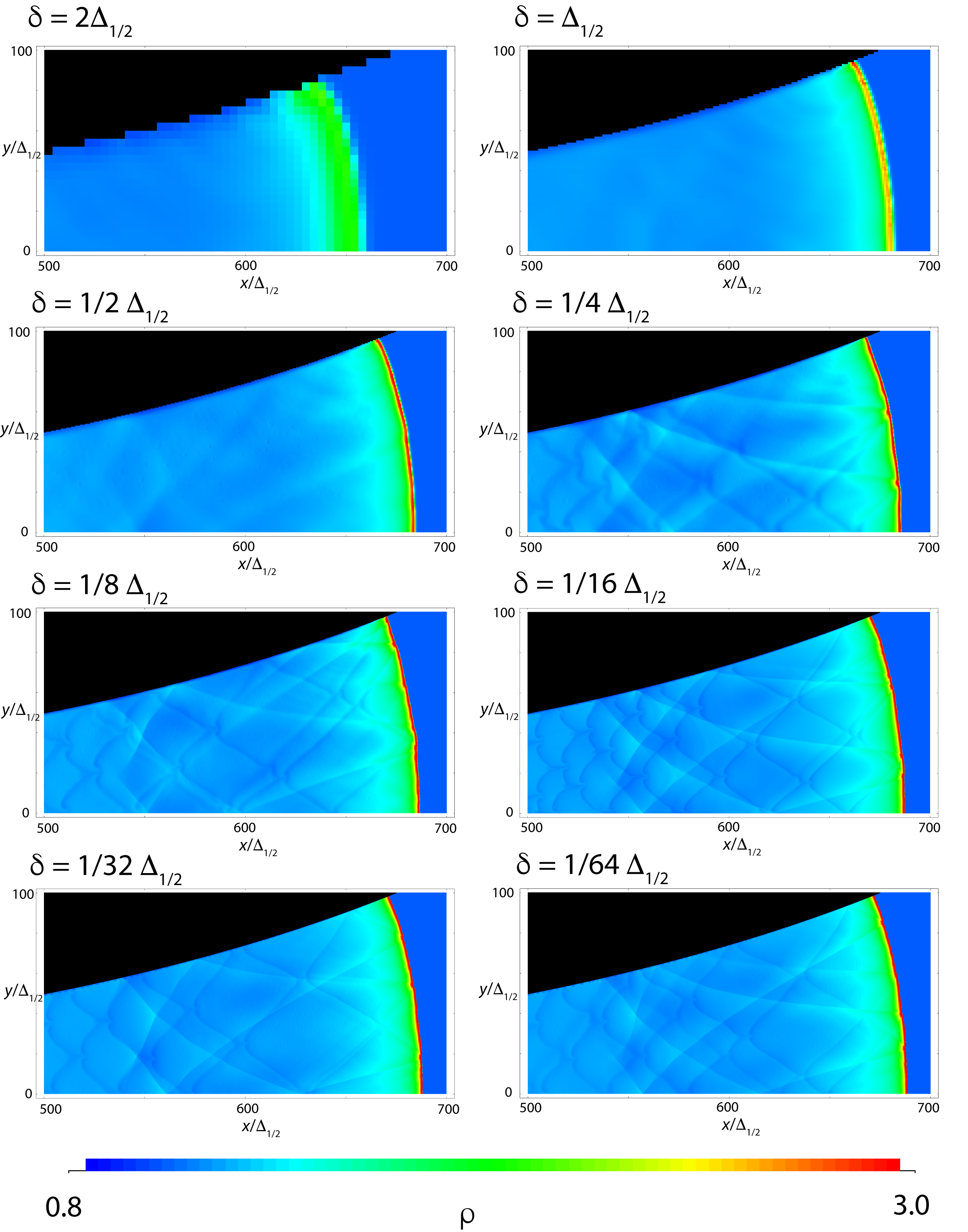}}
  \caption{Density field at $\bar{t}$=122 obtained with different minimum grid spacings $\delta$ for $\bar{K}$=0.004.}
\label{fig:K004resolution}
\end{figure}

Figure \ref{fig:K004resolution} shows the density fields obtained using the different spatial resolutions at $\bar{t}$=122, corresponding to the instant at which the detonation front has reached the end of the expanding channel (see also Fig. \ref{fig:K004_32pts}). With the exception of the least resolved simulation, the detonation front is approximately at the same location, and displays the same global characteristic curvature of the front. The cellular structure dynamics appear to converge approximately at a resolution of 1/16 $\Delta_{1/2}$.  The front cellular structure, the characteristic keystone features on its front behind the incident shock portions of the lead front and the dynamics of transverse shocks in the expanding gas behind the main reaction front are well resolved. 

\begin{figure}
  \centerline{\includegraphics[width=0.7\textwidth]{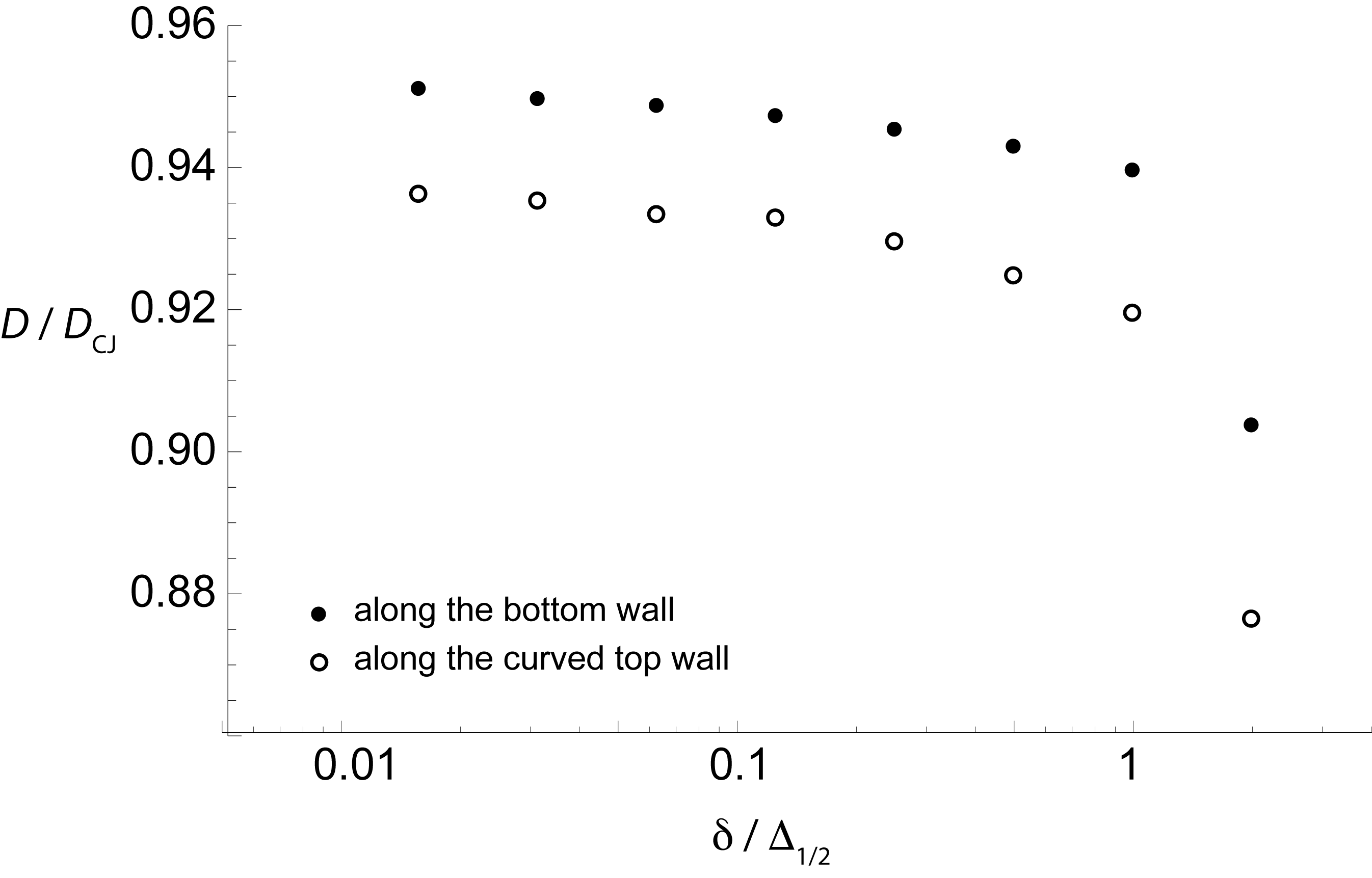}}
  \caption{Mean detonation front speeds between $\bar{t}$=42 and $\bar{t}$=122 along the axis and curved wall for different resolutions.}
\label{fig:K004resolutionspeeds}
\end{figure}

A more quantitative appraisal of numerical convergence was conducted on the detonation speed, since this is the eigenvalue for a given detonation front structure in the presence of lateral rate of strain.    For all the simulations conducted, the detonation speed was measured along the bottom straight wall and curved upper wall from $\bar{t}$=42 to $\bar{t}$=122.  The results obtained are shown in terms of the finest grid spacing $\delta$ in Fig.\ \ref{fig:K004resolutionspeeds}. The results indicate that a lower resolution tends to underestimate the detonation speed.  Between the two finest grid simulations conducted with 1/32 $\Delta_{1/2}$ and 1/64 $\Delta_{1/2}$, the speed difference is less than 0.002.  We thus conclude that this is approximately the error associated with the numerical determination of the speed eigenvalues for the multi-dimensional detonations reported in section 3.  

\bibliographystyle{jfm}
\bibliography{references}

\end{document}